\documentclass[%
 reprint, amsmath, amssymb, aps, superscriptaddress, prx]{revtex4-2}
\usepackage{graphicx}% Include figure files
\usepackage{bm}% bold math
\usepackage{siunitx}
\usepackage{soul} 
\usepackage{braket}
\usepackage{setspace}
\usepackage{bm, dsfont}% bold math
\usepackage[dvipsnames]{xcolor}

\usepackage[colorlinks, linkcolor=NavyBlue, citecolor=NavyBlue, urlcolor=NavyBlue, breaklinks=true]{hyperref}
\graphicspath{{Figures/}}
\begin{document}
\title{Quantum-enabled interface between microwave and telecom light}
\author{Rishabh Sahu}
\email{rsahu@ist.ac.at}
\author{William Hease}
\author{Alfredo Rueda}
\author{Georg Arnold}
\author{Liu Qiu}
\author{Johannes M. Fink}
\email{jfink@ist.ac.at}
\affiliation{Institute of Science and Technology Austria, am Campus 1, 3400 Klosterneuburg, Austria}%
\date{\today}
\begin{abstract}
Photons at telecom wavelength are the ideal choice for high density interconnects while solid state qubits in the microwave domain offer strong interactions for fast quantum logic.
Here we present a general purpose, quantum-enabled interface between itinerant microwave and optical light. We use a pulsed electro-optic transducer at millikelvin temperatures to demonstrate nanosecond timescale control of the converted complex mode amplitude with an input added noise of $N^{oe}_\textrm{in} = 0.16^{+0.02}_{-0.01}$ ($N^{eo}_\textrm{in} = 1.11^{+0.15}_{-0.07}$) quanta for the microwave-to-optics (reverse) direction. Operating with up to unity cooperativity, this work enters the regime of \textit{strong coupling cavity quantum electro-optics} characterized by unity internal efficiency and nonlinear effects such as the observed laser cooling of a 
superconducting cavity mode. The high quantum cooperativity of $C_q>10$ forms the basis for deterministic entanglement generation between superconducting circuits and light.
\end{abstract}
\maketitle

\vspace{0.1cm}
\noindent
\textbf{Introduction}\\
Superconducting qubits  
%\cite{devoret_superconducting_2013}
and semiconductor spin systems \cite{ClerkLehnertBertetEtAl2020} 
%Mi2018
are some of the most promising candidates for scalable quantum computing, but connecting many such microwave nodes to implement a scalable quantum network 
\cite{kimble_quantum_2008,wehner_quantum_2018}
remains a challenge \cite{Mangard2020}. Since higher energy optical photons are most suited to transfer quantum information via room temperature channels \cite{obrien_photonic_2009}, 
there has been a lot of development in the field of microwave to optical transduction \cite{lauk_perspectives_2020,lambert_quantum_2020}. 

A quantum-enabled transducer is characterized by an added equivalent input noise photon number of $N_\textrm{in}<1$ \cite{zeuthen_figures_2020}, a property that ideally is complemented with unitary scattering (bidirectional, near unity efficiency) and high bandwidth. These properties would be useful for various other purposes as well, such as for multiplexed control and readout of cryogenic electronics to facilitate the modular scaling requirements of current superconducting processors \cite{youssefi_cryogenic_2021,lecocq_control_2021}, for hybrid entanglement generation \cite{rueda_electro-optic_2019,Rueda2021} and quantum enhanced
remote sensing protocols \cite{barzanjeh_microwave_2015}, for sensitive radiometry \cite{bagci_optical_2014}, 
and multiplexed optical photon counting \cite{de_cea_photonic_2020}. 

Many hybrid devices using Rydberg atoms  
%\cite{vogt_efficient_2019}, 
magnons 
%\cite{hisatomi_bidirectional_2016,zhu_waveguide_2020} 
and rare-earth doped crystals 
%\cite{bartholomew_-chip_2020} 
have successfully demonstrated transduction between microwave and optical fields 
but with limited efficiency 
%but they have been limited to rather small efficiency 
and unspecified amounts of added input noise. Electro-opto-mechanical systems \cite{higginbotham_harnessing_2018,arnold_converting_2020} 
%andrews_bidirectional_2014
relying solely on the radiation pressure effect have shown record high conversion efficiencies of up to 47\% 
\cite{higginbotham_harnessing_2018} but owing to the low frequency mechanical mediator they are difficult to keep cool and thus far operate with kHz bandwidth in the classical limit.

Piezo-optomechanical devices \cite{Bochmann2013,forsch_microwave--optics_2020,mirhosseini_superconducting_2020}
%, jiang_efficient_2020,han_cavity_2020} 
% vainsencher_bi-directional_2016, Ramp2020, honl_microwave--optical_2021
exhibit higher bandwidths that match the typical requirements of superconducting qubit systems but face challenging impedance matching conditions, which so far limit the total conversion efficiency for itinerant fields. 
Electro-optic systems \cite{rueda_efficient_2016,Fan2018,hease_bidirectional_2020,holzgrafe_cavity_2020,mckenna_cryogenic_2020,witmer_siliconorganic_2020,fu_cavity_2021,xu_bidirectional_2020} 
%xu_bidirectional_2020  
also offer high bandwidth but typically require an even higher power optical pump to boost the conversion efficiency due to small vacuum coupling rates. In fully integrated systems this leads to optical heating and quasiparticle generation that adversely affects the superconducting part of the transducer.  

Recent experiments have used short pulses with low duty cycle to reduce the average thermal load
%to achieve higher conversion efficiencies while simultaneously maintaining a low thermal load 
\cite{forsch_microwave--optics_2020,mirhosseini_superconducting_2020,han_cavity_2020,fu_cavity_2021},
%xu_bidirectional_2020   
%Although short pulses with low duty cycle reduce the average thermal load, 
but this does not fully circumvent the instantaneous effects such as microwave cavity degradation due to Cooper pair breaking \cite{witmer_siliconorganic_2020} and detuning of the optical cavity due to the photo-refractive \cite{jiang_fast_2017} or the thermo-optic effect \cite{jiang_lithium_2019}.
Quasiparticles are especially detrimental for tightly integrated quantum systems \cite{mirhosseini_superconducting_2020}, as it prevents faithful optics-to-microwave conversion or the generation of deterministically entangled microwave-optical states. 

Here we present a modular electro-optic transducer that converts itinerant fields modes in both directions with $N_\textrm{in} \lesssim 1$. Since the microwave and optical cavity mode volumes are chosen to be large \cite{hease_bidirectional_2020}, we are able to apply high optical pump powers of up to \SI{500}{mW} without the aforementioned complications. This enables a total bidirectional conversion efficiency of $\eta_\textrm{tot}\approx 15\%$ and instantaneous values up to $30\%$ for preloaded cavity modes in the microwave-to-optics direction, i.e.~about two orders of magnitude higher than the previous record in the quantum limit \cite{mirhosseini_superconducting_2020}. Moreover, we demonstrate high precision phase coherent control of the spectral and temporal dynamics on nanosecond timescales and in excellent agreement with a comprehensive time-domain theory model that includes all five interacting field modes. 

\vspace{0.1cm}
\noindent
\textbf{Physics and implementation}\\
Electro-optic converters use the non-linear properties of a non-centrosymmetric crystal to couple microwave and optical fields \cite{tsang_cavity_2010,javerzac-galy_-chip_2016}.
%,tsang_cavity_2011}. 
We enhance the nonlinearity using a high quality cavity resonance for the optical and microwave fields \cite{ilchenko_whispering-gallery-mode_2003,rueda_efficient_2016,hease_bidirectional_2020}. For the optics, we use a whispering gallery mode resonator (WGMR) made from a 
z-cut lithium niobate wafer. The WGMR is enclosed in a machined 
aluminum cavity, which is designed to maximize the field overlaps and match the optical free spectral range (FSR).  

The total interaction Hamiltonian is given as $\hat{H}_\text{int} = \hbar g_0 (\hat{a}_e \hat{a}_p \hat{a}_o^\dagger + \hat{a}_e^\dagger \hat{a}_p \hat{a}_s^\dagger) + \textrm{h.c.}$, where, $g_0$ is the 
vacuum coupling rate and $\hat{a}_p$, and $\hat{a}_e$ represent the optical pump mode and microwave mode annihilation operators, respectively. $\hat{a}_o$ (optical signal mode), and $\hat{a}_s$ (Stokes mode) represent the optical mode annihilation operators on the blue and red side of the optical pump respectively. The first term in the Hamiltonian is the beam splitter interaction between modes $\hat{a}_e$ and $\hat{a}_o$ with the anti-Stokes scattering rate $\Gamma_{AS}$. The second term corresponds to the two-mode squeezing Hamiltonian between modes $\hat{a}_e$ and $\hat{a}_s$ with the Stokes scattering rate $\Gamma_{S}$.

\begin{figure}[t!]
	\centering
		\includegraphics[scale=0.83]{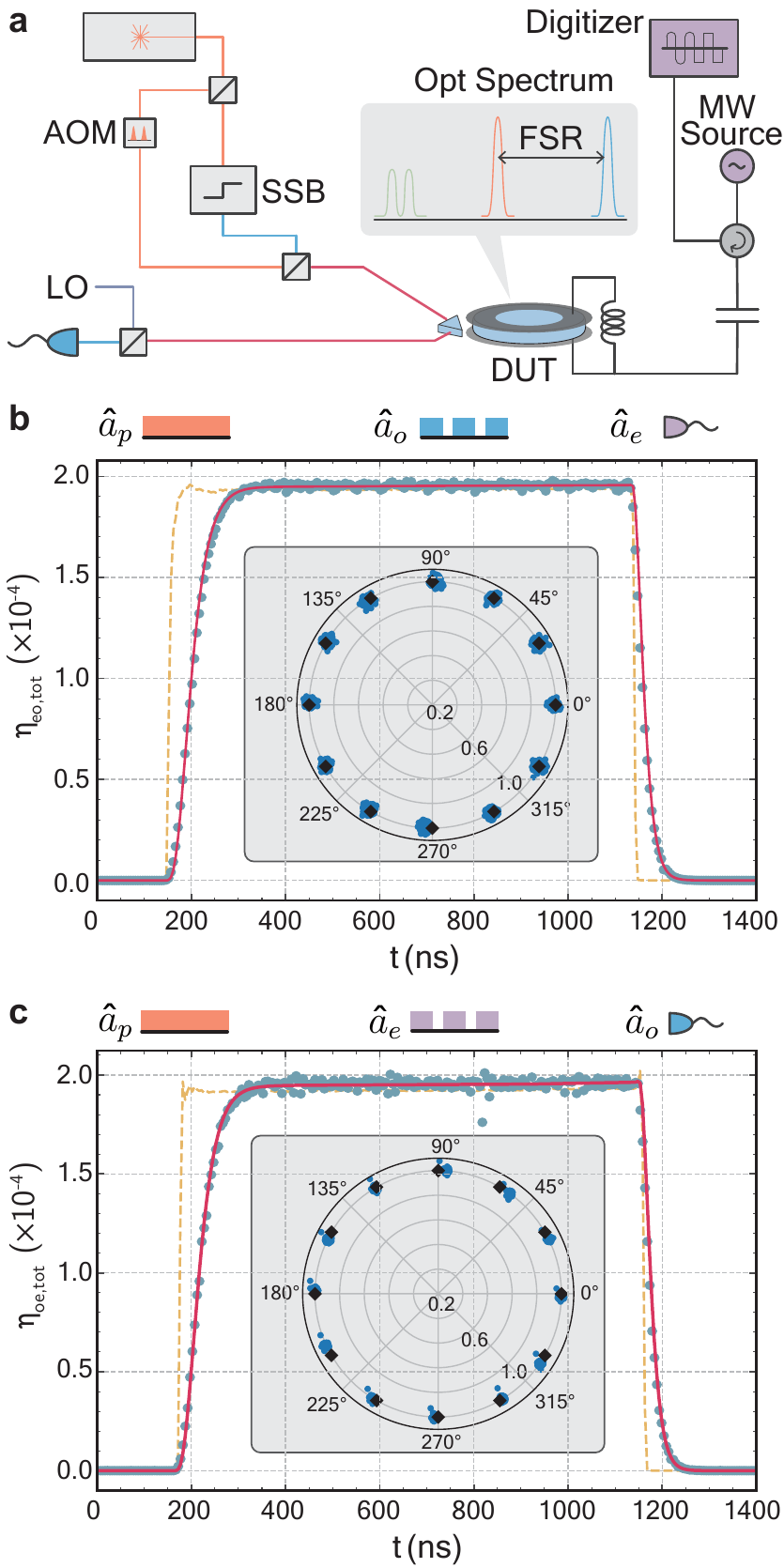}
	\caption{
	\textbf{Experimental schematic and characterization.
	}
\textbf{a}, Schematic of the experiment. The laser is divided in two parts - the optical pump and signal. The pump is pulsed via an acousto-optic modulator (AOM). The signal part is frequency up-shifted using a single sideband modulator (SSB). The optical pump and signal are combined and sent to the device under test (DUT) in dilution refrigerator via an optical fiber. The microwave signals are sent and received using a single port via a circulator. Optical and microwave output signals are measured with heterodyne detection. 
\textbf{b} (\textbf{c}), Optics-to-microwave (microwave-to-optics) conversion for a 
continuous wave optical pump 
of $P_p=\SI{134}{\mu W}$.
The measured input signal pulses are shown with dashed yellow lines and the converted signals with blue dots (red line is theory). 
Phase coherence and stability of the signal phase is shown in the insets. Rhombuses represent the phase imprinted on the input signal and the blue points represent phase values extracted from subsequently measured converted pulses.}
	\label{fig_1}
\end{figure}

The Stokes scattering causes phase-insensitive amplification adding extra noise to the process of transduction. We suppress this amplification through cross-polarization coupling of the Stokes mode $\hat{a}_s$ with a polarization-orthogonal  degenerate optical mode, $\hat{a}_r$ \cite{rueda_efficient_2016,werner_control_2018}. 
This hybridizes the optical mode $\hat{a}_s$ limiting its participation in the dynamics, as schematically shown in the optical spectrum in Fig.~\ref{fig_1}(a). The full Hamiltonian is then given as
\begin{equation}
\hat{H}_\text{int} = \hbar g_0 (\hat{a}_e \hat{a}_p \hat{a}_o^\dagger + \hat{a}_e^\dagger \hat{a}_p \hat{a}_s^\dagger) + i J \hat{a}_s^\dagger \hat{a}_r + \textrm{h.c.},
\label{eq:hamiltonian}
\end{equation}
where $J$ is the cross-polarization coupling rate between the $\hat{a}_s$ and $\hat{a}_r$ modes.

The figure of merit in electro-optic systems is the cooperativity $C=4\bar{n}_pg_0^2/(\kappa_o\kappa_e)$, where $\bar{n}_p$ stands for the number of optical pump photons inside the optical resonator and  $\kappa_{o}$ ($\kappa_{e}$) for the total loss rate of the participating optical (microwave) mode. In analogy, we define the cooperativity for characterizing the mode coupling $C_J = 4J^2/(\kappa_s \kappa_r)$. On resonance, the total transduction efficiency including the gain due to amplification is calculated in terms of the two cooperativities as
\begin{equation}
\eta_{tot} = \eta_o \eta_e \Lambda^2\frac{4C(1+C_J^{-1})^2}{(1+C+C_J^{-1})^2},
\label{eq:conv_eff}
\end{equation}
with the two mode coupling efficiencies $\eta_i= \kappa_{i,\textrm{ex}}/\kappa_{i}$ with $i=e,o$ and the factor $\Lambda$ accounting for optical coupling loss due to spatial mode-mismatch between the Gaussian input beam and the optical mode's evanescent field. 
The gain due to amplification depends on the suppression ratio $\mathcal{S} = \Gamma_{S}/\Gamma_{AS} = (1 + C_J)^{-1}$. For $J\rightarrow\infty$, i.e. perfect suppression $\mathcal{S}=0$, Eq.~\ref{eq:conv_eff} reduces to the simple triply resonant model, as shown in the Supplementary Information.

A simplified schematic of the experiment is shown in Fig.~\ref{fig_1}(a). Here we use the same device as in Ref.~\cite{hease_bidirectional_2020} with a frequency tunable 
microwave mode with internal linewidth $\kappa_{e,\textrm{in}}/(2\pi )=$ \SI{8.1}{MHz}, 
coupling efficiency $\eta_e=0.41$ 
and mode frequency $\omega_e/(2\pi )= $ \SI{8.795}{GHz} that accurately matches the optical FSR.
The optical WGMR supports high quality modes with an internal linewidth of $\kappa_{o,\textrm{in}}/(2\pi )=$ \SI{10.8}{MHz}. 
The optical input and output is coupled to the WGMR mode via frustrated total internal reflection through an in-situ movable diamond prism with 
$\Lambda\approx0.78$ and 
$\eta_o = 0.58$ in this experiment. 
The optical mode $\hat{a}_r$ has a total linewidth of $\kappa_r/(2\pi )=$ \SI{11.7}{MHz} and is coupled to mode $\hat{a}_s$ with a coupling rate of $J/(2\pi )=$  \SI{27}{MHz}. 

These values are obtained from a set of characterization measurements 
including time-domain scattering parameter measurements.
Figure \ref{fig_1}(b) and (c) shows 
the calibrated time dependent measurement of a converted signal pulse in case of microwave-to-optics and optics-to-microwave conversion, respectively. These pulses are measured with a bandwidth of \SI{200}{MHz} and shown together with a fit to the numerical model (red line) that takes the measured input pulses (yellow dashed lines) with a rise time of \SI{15}{ns} and \SI{5}{ns} respectively as input data. We find excellent agreement of all four time dependent scattering parameters and other independent characterization measurements,
see Supplementary Information. 
The 10\% to 90\% rise time of the converted pulses in both directions of \SI{85}{ns} is limited by the linewidths of the optical and microwave modes in this case of comparably low cooperativity $C=$ \num{3.4E-4}. Moreover, we
also explicitly verify the faithful phase control and stability over subsequent pulses in both directions as shown in the insets.

\begin{figure*}[t!]
	\centering
		\includegraphics[width=\textwidth]{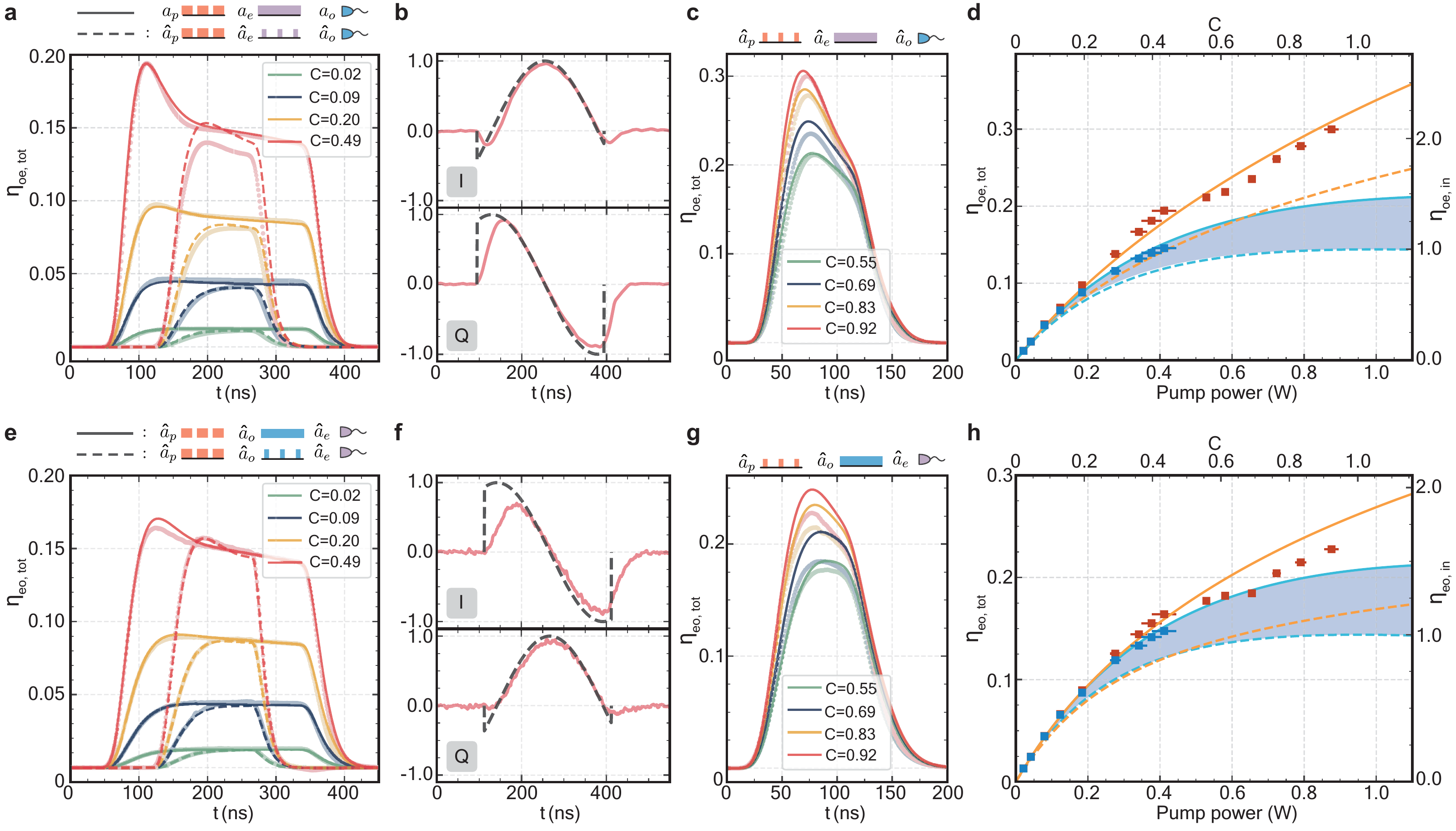}
	\caption{\textbf{High cooperativity bidirectional conversion. }
Top (bottom) row panels show results for microwave-to-optics (optics-to-microwave) conversion.
\textbf{a} (\textbf{e}), Converted signal pulses (bright dots are calibrated measurements and lines are theory) for a \SI{300}{ns} long optical pump pulse measured with a \SI{500}{Hz} repetition rate for  
a CW signal (solid lines) and a pulsed signal (dashed lines). 	
\textbf{b} (\textbf{f}), Measured IQ quadrature modulation of the converted signal pulse (red lines) and the applied IQ quadrature modulation to the input signal (gray dashed lines) for $C=0.49$. 
\textbf{c} (\textbf{g}), Converted signal pulses (bright dots are calibrated measurements and lines are theory) for a \SI{100}{ns} long optical pump pulse and a CW signal measured with a \SI{500}{Hz} repetition rate.  
\textbf{d} (\textbf{h}), Summary of measured steady state (blue squares) and peak (red squares) total (left axis) and internal (right axis) conversion efficiency as a function of pump power and cooperativity. 
Blue lines show the theory prediction for the steady state efficiency (Eq.~\ref{eq:conv_eff}) and orange  lines  for the peak efficiency (see Supplementary Information), respectively.
Solid lines take into account the suppression ratio $\mathcal{S}= 0.22$ while dashed lines are for the ideal case $\mathcal{S}=0$. 
The blue shaded area represents the gain of the steady state transduction efficiency due to finite sideband suppression. The vertical (smaller than the symbols) and horizontal error bars represent the standard error of the individual measured time averaged values.
} 
	\label{fig_2}
\end{figure*}

\vspace{0.1cm}
\noindent
\textbf{High-cooperativity conversion}\\
Higher efficiency can be achieved with higher $C$ as a result of higher pump power. To keep the thermal load low, we pulse the optical pump with a \SI{500}{Hz} repetition rate. Figure \ref{fig_2}(a) and (e) shows the calibrated efficiency for microwave-to-optics and optics-to-microwave conversion respectively for different cooperativities. 
The converted pulses are measured (\SI{200}{MHz} bandwidth) for two cases, i.e. a CW signal (solid lines) and a pulsed signal while the optical pump pulse is on (dashed lines). 
The solid and dashed lines show the theoretical prediction from the numerical model 
with the input optical loss as the only fit parameter. We generally find excellent agreement and assign the observed mismatch in Fig.~\ref{fig_2}(a) for the case of pulsed signals to a small amount of uncorrected output filter drift. 

We observe a large overshoot in the beginning of the converted pulse for CW signals (solid lines) and a smaller one for high cooperativity pulsed signals (dashed lines). The former appears because the cavity is pre-loaded with photons which get converted immediately when the optical pump pulse arrives. 
This momentarily bypasses the coupling loss to the cavity
and since $\eta_o\ne\eta_e$, the magnitude of this overshoot is different for the two directions of conversion. Importantly, there is also an overshoot for the case of pulsed signals (dashed lines) with $C\gtrsim 0.5$ because at high cooperativity the coherent interaction rate approaches that of the two loss rates. This feature therefore represents the onset of coherent oscillations between microwave and optical photons in a strongly coupled electro-optic system.

For the measurement with $C=\num{0.49}$ we explicitly show real-time phase control in Fig.~\ref{fig_2}(b) and (f). 
A linear phase change is imprinted on the input signal pulses via IQ quadrature modulation while keeping the amplitude constant. The I and Q quadrature of signal pulse are, thus, sine and cosine functions respectively with frequency \SI{2}{MHz}. The measurements of the two converted quadratures match the input modulation closely. The only exception is during the beginning and the end of the optical pump pulse due to finite transducer bandwidth of 18 MHz for this $C$.  

In order to reach $C\approx1$ without entering an observed regime of instability, see Supplementary Information, 
we use shorter, \SI{100}{ns} long optical pump pulses as shown in Fig.~\ref{fig_2}(c) and (g).
The highest microwave-to-optical conversion efficiency momentarily reaches up to \num{30}\% for a 
CW signal tone. In very good agreement with theory, this is a result of three effects, i.e.~the preloading of the microwave cavity, the increased gain for higher $C$, and the strong and coherent electro-optic interaction, which would reveal an oscillatory behavior if longer pulses could be sustained. The observed deviations from theory in Fig.~\ref{fig_2}(g) are caused by the slight broadening of the microwave mode linewidth due to the increased average bath temperature, in agreement with CW pump experiments \cite{hease_bidirectional_2020}, an effect which the theory model does not take into account. 

A summary of conversion efficiencies as a function of optical pump power and corresponding cooperativity is presented in Fig.~\ref{fig_2}(d) and (h). The red and blue colors show peak and steady-state conversion efficiencies respectively. Solid lines are predicted conversion efficiencies for our experimental parameters corresponding to $\mathcal{S}= 0.22$, while dashed lines represent the case of perfect suppression $\mathcal{S}=0$. The blue shaded area thus represents the gain due to the not fully suppressed Stokes process.

\vspace{0.1cm}
\noindent
\textbf{Thermal and quantum noise}\\
High conversion efficiencies are an expected outcome of high optical pump powers. To be useful for quantum applications, they must be accompanied with low added noise. Since our microwave cavity is machined from a bulk aluminum piece, we do not observe indications of added noise due to quasi-particle generation. Instead, we have two main sources of noise - thermal noise due to optical pump heating and amplified vacuum noise due to finite gain. 
We find that the thermal heating is a function of the average optical pump power $P_\text{avg}$ which depends on the pulse length and duty cycle, as shown in the Supplementary Information. In contrast, the amplified vacuum noise depends only on the instantaneous power of the pulse $P_p$, which determines the cooperativity. 

\begin{figure*}[t!]
	\centering
		\includegraphics[width=0.88\textwidth]{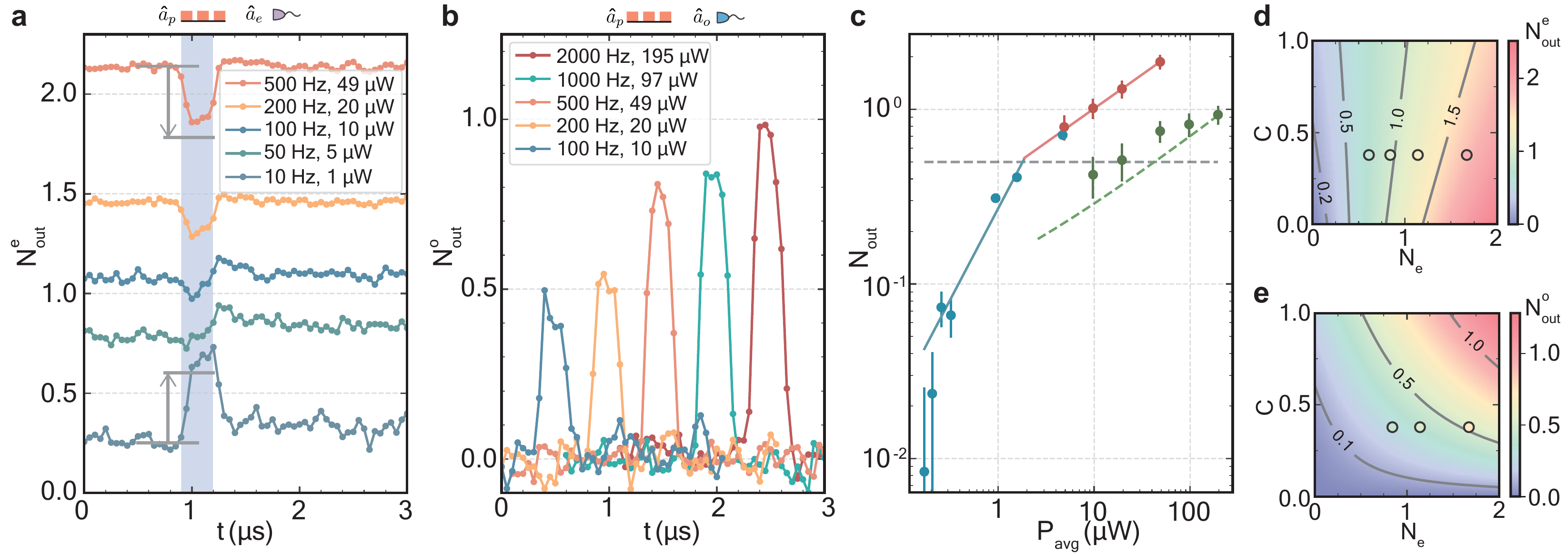}
	\caption{\textbf{Measured added noise during conversion. }
\textbf{a}, $N^{e}_{\textrm{out}}$ for a \SI{300}{ns} long optical pump pulse ($C= 0.38$) for different repetition rates (average optical pump powers) measured with a \SI{10}{MHz} bandwidth centered at the microwave resonance. The blue shaded region shows the times span when the optical pump is on. 
\textbf{b}, $N^{o}_{\textrm{out}}$ for the same optical pump pulses and various repetition rates (average optical pump powers) measured with a \SI{10}{MHz} bandwidth centered at the optical signal frequency. 
\textbf{c}, Compilation of all measured $N_{\textrm{out}}$ for different average optical pump powers. Red and blue points represent measured $N^{e}_{\textrm{out}}$ with \SI{10}{MHz} and \SI{100}{kHz} bandwidths respectively. Green points represent $N^{o}_{\textrm{out}}$ measured with \SI{10}{MHz} bandwidth. The solid lines represent power law fits. The green dashed line shows the predicted $N^{o}_{\textrm{out}}$ using the measured $N^{e}_{\textrm{out}}$. All the points are taken with $C =$\num{0.38} ($\eta_{\textrm{tot}}=11.4$\%) except for the first three blue points which correspond to $C=$\num{0.20} ($\eta_{\textrm{tot}}=8.0$\%), \num{0.24} ($\eta_{\textrm{tot}}=9.1$\%), and \num{0.3} ($\eta_{\textrm{tot}}=10.3$\%) respectively. The error bars are inferred as in Ref.~\cite{hease_bidirectional_2020}. For low occupancies the confidence interval is given by the systematic error due to the absolute drift of the measurement baseline of $\pm 0.02$ photons.
\textbf{d}, Predicted $N^{e}_{\textrm{out}}$ (\SI{10}{MHz} bandwidth) as a function of $C$ and $N_e$ for $\mathcal{S}= 0.22$. Relevant measurements from (a) are shown as circles where experimental values are represented by the inside color. 
\textbf{e}, Predicted $N^{o}_{\textrm{out}}$ (\SI{10}{MHz} bandwidth) as a function of $C$ and $N_e$. The lowest three measurements from (b) are shown as circles where experimental values are represented by the inside color.}
	\label{fig_3}
\end{figure*}

Figure~\ref{fig_3}(a) captures the effect of both types of added noise. Here we show the measured microwave noise output $N^{e}_{\textrm{out}}$ in the time domain with a \SI{10}{MHz} filter centered at the microwave resonance, when \SI{300}{ns} long optical pump pulses with $C= 0.38$ are applied (shaded region) with different repetition rates from 10 to 500 Hz. High repetition rates increase the measured average thermal output noise, which stays approximately constant during the measurement time of \SI{3}{\mu s}. During the pulse however, we observe either a classical or a quantum effect depending on the average thermal occupancy of the microwave mode. For higher mode temperature (red curve), parametric laser cooling of the microwave mode \cite{tsang_cavity_2010},
due to up-conversion of noise to optics is dominant and in agreement with theory (gray arrow). But as the thermal noise is decreased for the same cooperativity, additional noise due to vacuum amplification overwhelms the parametric cooling effect. For the lowest occupancy curve with a \SI{10}{Hz} repetition rate, the vacuum amplification is clearly observed during the pump pulse and in good agreement with theory (gray arrow). This last curve is measured with a lower suppression $\mathcal{S}\approx 0.82$ in order to enhance the effect for a better signal to noise ratio. We assign slight mismatches between the theoretical prediction and experiment to an expected small amount of thermal heating during the pulse.

The observation of parametric laser cooling implies the presence of noise at the output of the optical signal mode $\hat{a}_o$. $N^{o}_{\textrm{out}}$ is measured with a \SI{10}{MHz} bandwidth around the optical signal frequency and shown in Fig.~\ref{fig_3}(b). The optical pump pulses are the same as in Fig.~\ref{fig_3}(a) with varying repetition rates. With higher 
$P_\text{avg}$, the thermal microwave mode occupancy increases, thus, increasing the output optical noise. 

We summarize these results in Fig.~\ref{fig_3}(c) as a function of $P_\text{avg}$. The $N_{\textrm{out}}$ during the pulse from Fig.~\ref{fig_3}(a) are shown as red points, and from Fig. \ref{fig_3}(b) as green points, along with corresponding power law fits (solid lines). The dashed green line represents the predicted optical noise based on theory and the fitted microwave noise. In addition, we show microwave noise measurements conducted with a \SI{100}{kHz} bandwidth for better signal to noise ratio at the lowest occupancies where the system is deep in its quantum ground state (blue points), in agreement with Ref.~\cite{hease_bidirectional_2020}. 

In Fig.~\ref{fig_3}(d) and (e) we show the theoretical prediction of the $N^{e}_{\textrm{out}}$ and $N^{o}_{\textrm{out}}$ respectively as a function of the steady state microwave mode occupancy $N_e$ and cooperativity $C$ for the relevant experimental case $\mathcal{S}= 0.22$. 
As a function of $N_e$, in Fig.~\ref{fig_3}(d) the contours change from left leaning to right leaning as the transition from quantum amplification to classical cooling occurs. The 
measurements from Fig.~\ref{fig_3}(a) are shown as circles where the inside color represents the experimentally measured value in excellent agreement with the theoretical prediction. Similarly, the predicted dependence of $N^{o}_{\textrm{out}}$ is shown in Fig.~\ref{fig_3}(e) where we included the
measurements from Fig.~\ref{fig_3}(b) with good agreement with theory. 
These results provide strong evidence that the noise mechanisms are understood in detail and that it is predominantly $P_\text{avg}$ that determines the thermal noise limitations of this device.

\begin{figure}[t!]
	\centering
		\includegraphics[width=0.90\columnwidth]{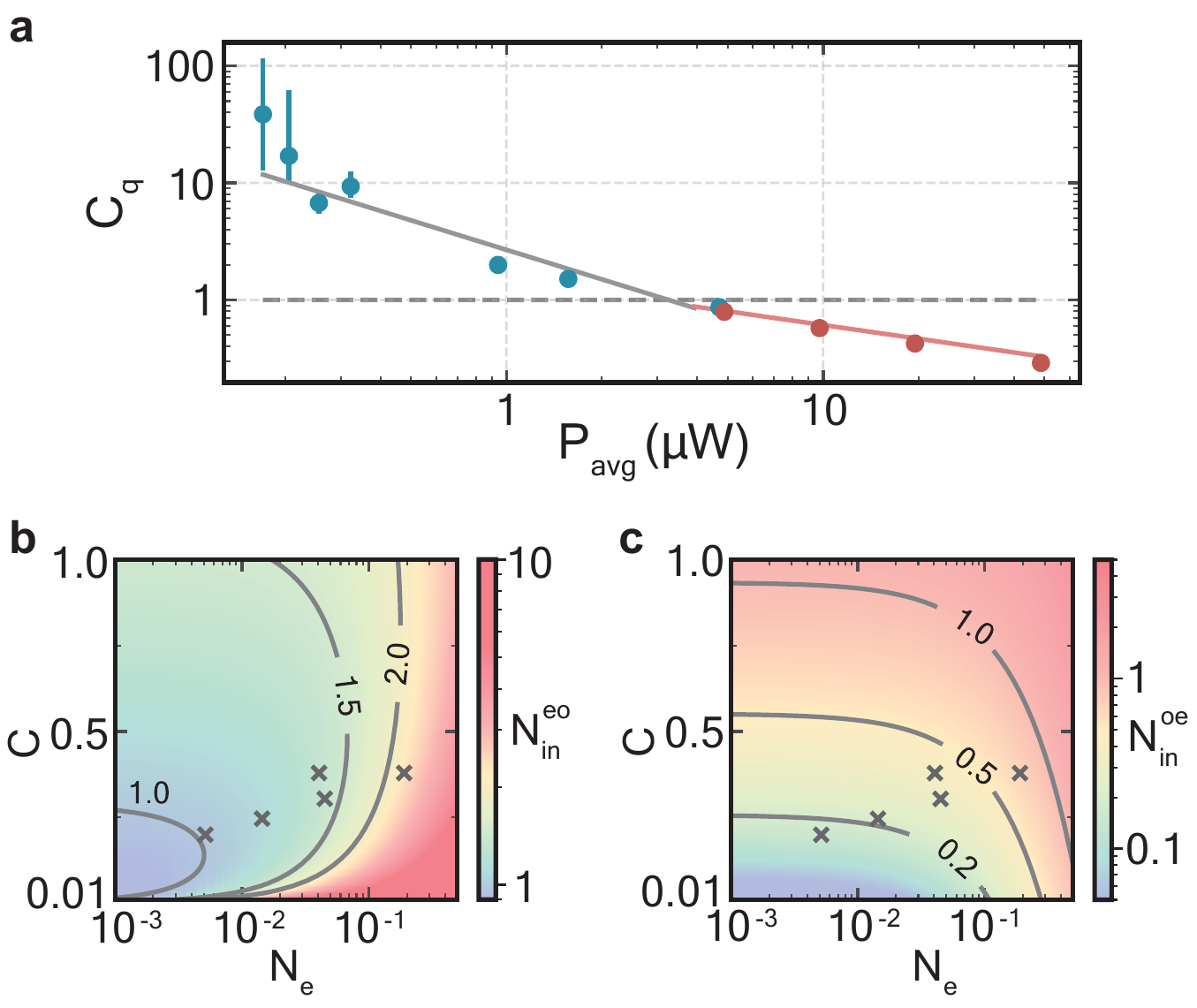}
	\caption{
	\textbf{Quantum cooperativity and equivalent input added noise.}  
	\textbf{a}, Inferred quantum cooperativity $C_q = C/N_e$ for the measured output noise values shown in Fig.~\ref{fig_3}(c). 
	\textbf{b} (\textbf{c}), Inferred optics-to-microwave $N^{eo}_\textrm{in}$ (microwave-to-optics $N^{oe}_\textrm{in}$) equivalent added input noise as a function of $C$ and $N_e$ for $\mathcal{S}= 0.22$. Crosses mark the blue measurement points in Fig.~\ref{fig_3}(c).}
	\label{fig_4}
\end{figure}

Low occupancies and high cooperativity are the preconditions for interesting cavity quantum electro-optics experiments in analogy to cavity optomechanics \cite{tsang_cavity_2010} as well as for quantum limited conversion. Figure \ref{fig_4}(a) shows the calculated quantum cooperativity $C_q = C/N_e$ for the measurements in Fig.~\ref{fig_3}(c) where $N_e$ is inferred from the measured $N_\textrm{out}^{e}$. With $C_q\gg1$ our device can not only be used for quantum-enabled conversion but also for two mode squeezing of hybrid microwave and optical field states \cite{rueda_electro-optic_2019}. 

Using $N_e$ and $C$, we can also infer the $N_\textrm{out}$ that includes both thermal and quantum noise contributions and calculate the resulting equivalent added noise referenced to the input of the converter $N_\textrm{in} = N_\textrm{out}/\eta_\textrm{tot}$.
Fig.~\ref{fig_4}(b) and (d) shows the equivalent input added noise for optics-to-microwave ($N^{eo}_\textrm{in}$) and microwave-to-optics ($N^{oe}_\textrm{in}$) conversion respectively as a function of $N_e$ and $C$ for $\mathcal{S}= 0.22$. The crosses mark the measured parameters we achieved experimentally. For $N^{eo}_\textrm{in}$, we reach $1.11^{+0.15}_{-0.07}$, and for $N^{oe}_\textrm{in}$, we reach as low as $0.16^{+0.02}_{-0.01}$ equivalent added noise photons. Here the confidence interval is taken from error propagation using the confidence interval of $N_\textrm{out}$.

\vspace{0.1cm}
\noindent
\textbf{Conclusion and Outlook}\\
In this work we have demonstrated a highly efficient mode converter with nanosecond timescale phase and amplitude control interfacing the microwave X with the telecom C band. With its high bandwidth of up to 24 MHz, an equivalent input noise $N^{oe}_\textrm{in}\ll1$ and a quantum cooperativity $C_q\gg1$ it is suitable for faithful quantum transduction tasks and ideal for optically heralded superconducting quantum networking schemes \cite{krastanov_optically-heralded_2020}. 
Other examples include the
generation of narrow band Fock states at telecom wavelength via transduction from on-demand microwave single photon sources \cite{reed_faithful_2017,Mangard2020} and the generation and verification of superconducting qubit - optical photon entanglement. 

The new parameter regime of high cooperativity electro-optics also enables first systematic studies of dynamical and quantum back-action in electro-optic systems \cite{tsang_cavity_2010} as well as the deterministic generation of microwave-optics entanglement in the continuous variable domain \cite{rueda_electro-optic_2019}. Since we reach very low $N^{oe}_\textrm{in}$ - comparable to near quantum limited Josephson parametric amplifiers - such devices could furthermore also become interesting for fiber-based multiplexed optical single-shot readout of large scale superconducting processors \cite{youssefi_cryogenic_2021}. 
 
Current limitations are the somewhat complex assembly with free space optical coupling, the need for high pump powers which leads to relatively slow pulse repetition rates, and the finite suppression ratio of the Stokes process which is the dominant source of added noise at low average pump power. These challenges can be addressed with more robust and compact packaging such as the proposed center-clamped design \cite{rueda_electro-optic_2019} that should also reduce phonon losses that might currently limit the microwave mode quality factor. Fabrication and material improvements have already led to WGMR quality factors that are close to 2 orders of magnitude higher than reported here. 
Maintaining this in a full device at low temperature would reduce the required pump power by four orders of magnitude for the same $C$, thus significantly reducing the thermal load and allowing for much faster repetition rates. Narrower optical linewidths potentially also help to further improve the suppression of the Stokes process and the resulting amplified vacuum noise.
Higher internal quality factors will furthermore be key to achieve higher external mode coupling ratios with the same bandwidth - a necessary requirement to push the total itinerant conversion efficiency towards unity. 

The data and code used to produce the figures in this manuscript will be made available on Zenodo. 

\vspace{0.1cm}
\noindent
\textbf{Acknowledgments}\\
The authors thank S. Wald and F. Diorico for their help with optical filtering, O. Hosten and M. Aspelmeyer for equipment, H.G.L. Schwefel for materials and discussions, L. Drmic and P. Zielinski for software support and the MIBA workshop at IST Austria for machining the microwave cavity.
This work was supported by the European Research Council under grant agreement No 758053 (ERC StG QUNNECT) and the European Union's Horizon 2020 research and innovation program under grant agreement No 899354 (FETopen SuperQuLAN). W.H. is the recipient of an ISTplus postdoctoral fellowship with funding from the European Union's Horizon 2020 research and innovation program under the Marie Sk\l{}odowska-Curie grant agreement No. 754411. G.A. is the recipient of a DOC fellowship of the Austrian Academy of Sciences at IST Austria. J.M.F. acknowledges support from the Austrian Science Fund (FWF) through BeyondC (F7105) and the European Union's Horizon 2020 research and innovation programs under grant agreement No 862644 (FETopen QUARTET).

\vspace{0.1cm}
\noindent
\textbf{Author contributions} \\
R.S., W.H. and A.R. worked on the setup and performed the measurements. R.S. did the data analysis. R.S., A.R. and L.Q. developed the theory. G.A. helped with the heterodyne calibration. R.S. and J.M.F wrote the manuscript with contributions from all authors. J.M.F. supervised the project. 

%%%%%%%%%%%%%%%%%%%%%%%%%%%%%%%%
%%%%%%%%%%%%%%%%%%%%%%%%%%%%%%%%
%\twocolumngrid
\bibliography{RiSa_EO_paper_pulsed}
%%%%%%%%%%%%%%%%%%%%%%%%%%%%%%%%
%%%%%%%%%%%%%%%%%%%%%%%%%%%%%%%%
\onecolumngrid
\newpage
%\begin{center}
%\Large
%Supplementary Material 
%\end{center}
%\maketitle
\appendix
\tableofcontents
%%%%%%%%%%%%%%%%%%%%%%%%%%%%%%%%
%%%%%%%%%%%%%%%%%%%%%%%%%%%%%%%%
\newpage
\section{Theory}
\label{sec:theory}
Our quantum transducer uses the $\chi^{(2)}$ nonlinearity in Lithium Niobate to directly couple microwave and optical fields \cite{ilchenko_whispering-gallery-mode_2003,tsang_cavity_2010}. The nonlinearity is enhanced by utilizing a cavity resonance for all involved modes. A high quality whispering gallery mode resonator (WGMR) supports the optical modes. It is placed inside a circular microwave cavity whose $m=1$ mode is in-situ frequency tunable and accurately matched with the optical FSR. We use the highest $r_{33}$ electro-optic coefficient to couple the transverse electric (TE) polarized microwave and optical fields. 
%Consequently, both the microwave and optical fields are transverse electric (TE) polarized. 

\subsection{Hamiltonian}
The microwave mode couples with the optical pump and the two optical sidebands on either side of the optical pump. The interaction Hamiltonian is given as,
\begin{equation}
\hat{H}_\mathrm{int} = \hbar g_0 (\hat{a}_e \hat{a}_p \hat{a}_o^\dagger + \hat{a}_e^\dagger \hat{a}_p \hat{a}_s^\dagger) + \textrm{h.c.},
\label{eq:h_int_0}
\end{equation}
where $g_0$ is the non-linear vacuum coupling strength, and $\hat{a}_p$, and $\hat{a}_e$  represent optical pump mode and microwave mode annihilation operators respectively. $\hat{a}_o$ and $\hat{a}_s$ represent the optical sideband mode annihilation operators on the blue (optical signal mode) and the red (optical Stokes mode) side of the optical pump separated by one optical FSR. As noted in the main text, the first term in Eq.~\ref{eq:h_int_0} represents a beam splitter interaction between the microwave mode $\hat{a}_e$ and the optical signal mode $\hat{a}_o$ as needed for noiseless conversion between microwave and optics fields. The second term is a two mode squeezing interaction due to interaction of the microwave mode $\hat{a}_e$ and the optical Stokes mode $\hat{a}_s$ also mediated by the optical pump mode. This term is responsible for phase insensitive amplification, which adds extra noise in the process of transduction and should be suppressed in an ideal transducer.

\begin{figure}[t!]
	\centering
		\includegraphics[scale=1.5]{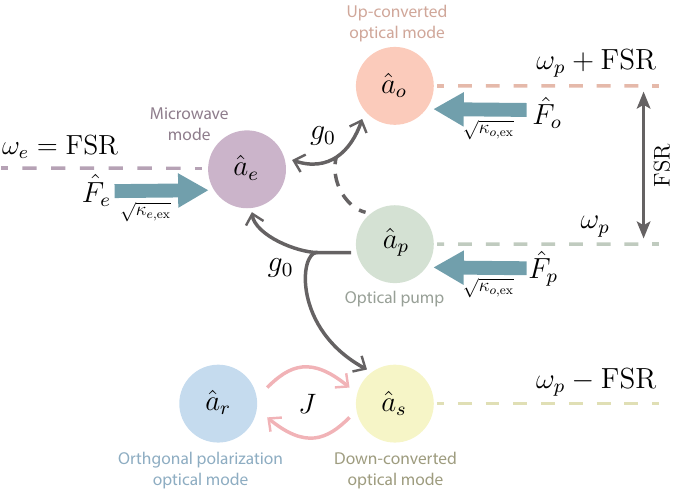}
	\caption{\textbf{Participating modes and their interactions.} In the presented converter, there are three optical TE modes - the optical pump mode $\hat{a}_p$ and one optical mode on the blue ($\hat{a}_o$) and red  ($\hat{a}_s$) side of the optical pump separated by one FSR each. The optical mode $\hat{a}_o$ is parametrically coupled to the microwave mode $\hat{a}_e$ via the beam splitter interaction with the vacuum coupling rate $g_0$. The microwave mode $\hat{a}_e$ and the optical mode $\hat{a}_s$ also stimulate the optical pump $\hat{a}_p$ to down-convert into $\hat{a}_s$ and $\hat{a}_e$ via the two-mode squeezing interaction. This two-mode squeezing interaction is suppressed by coupling the optical mode $\hat{a}_s$ to a degenerate TM optical mode $\hat{a}_r$ with coupling strength $J$ creating an anti-crossing at the $\hat{a}_s$ mode frequency. 
%The modes $\hat{a}_s$ and $\hat{a}_r$ are coupled with coupled rate $J$. 
In our experiment, we use three coherent inputs - $\hat{F}_p$ for the optical pump, $\hat{F}_e$ for the microwave mode, and $\hat{F}_o$ for the optical signal mode to probe the system dynamics.}
	\label{fig_0}
\end{figure}

In our experiment, we suppress the participation of the optical mode $\hat{a}_s$ by coupling this mode to a degenerate transverse magnetic (TM) optical mode $\hat{a}_r$, as shown in Fig.~\ref{fig_0}. This creates an avoided crossing in the TE mode $\hat{a}_s$ \cite{rueda_efficient_2016,werner_control_2018}. The pump then interacts with this hybridized $\hat{a}_s$ mode reducing the density of states in the relevant frequency range. The full interaction Hamiltonian for our system is
\begin{equation}
\hat{H}_\mathrm{int} = \hbar g_0 (\hat{a}_e \hat{a}_p \hat{a}_o^\dagger + \hat{a}_e^\dagger \hat{a}_p \hat{a}_s^\dagger) + i J \hat{a}_s^\dagger \hat{a}_r + \textrm{h.c.},
\end{equation}
where $J$ is the coupling rate between the $\hat{a}_s$ and the $\hat{a}_r$ modes. 

\subsection{Equations of motion}
We derive the time dynamics of these operators using the Heisenberg equations of motion. The resulting system of equations is the following:
\begin{subequations}
\label{eq:timederivatives2} 
\begin{eqnarray}
& \dot{\hat{a}}_p = -i \Delta_p \hat{a}_p -\frac{\kappa_o}{2} \hat{a}_p   -ig_0 (\hat{a}_s\hat{a}_e + \hat{a}_e^\dagger \hat{a}_o) + \Lambda\sqrt{\kappa_{o,\textrm{ex}}}\bar{F}_p + \sqrt{\kappa_{o,\textrm{ex}}}\delta\hat{a}_{p,e} + \sqrt{\kappa_{o,\textrm{in}}}\delta\hat{a}_{p,\textrm{in}} ,\label{eq:pump_eqn}
\\
& \dot{\hat{a}}_o = -i \Delta_o \hat{a}_o -\frac{\kappa_o}{2} \hat{a}_o -ig_0 \hat{a}_p\hat{a}_e  + \Lambda\sqrt{\kappa_{o,\textrm{ex}}}\bar{F}_o + \sqrt{\kappa_{o,\textrm{ex}}}\delta\hat{a}_{o,\textrm{ex}} + \sqrt{\kappa_{o,\textrm{in}}}\delta\hat{a}_{o,\textrm{in}},\label{eq:plus_fsr_eqn}
\\
& \dot{\hat{a}}_s = -i \Delta_s \hat{a}_s -\frac{\kappa_o}{2} \hat{a}_s -ig_0 \hat{a}_p\hat{a}_e^\dagger  - i J \hat{a}_r + \sqrt{\kappa_{o,e}}\delta\hat{a}_{s,\textrm{ex}} + \sqrt{\kappa_{o,\textrm{in}}}\delta\hat{a}_{s,\textrm{in}}  ,\label{eq:minus_fsr_eqn}
\\
& \dot{\hat{a}}_e = -i \Delta_e \hat{a}_e -\frac{\kappa_e}{2} \hat{a}_e -ig_0 \hat{a}_p\hat{a}_s^\dagger -ig_0 \hat{a}_p^\dagger \hat{a}_o  + \sqrt{\kappa_{e,\textrm{ex}}}\bar{F}_e + \sqrt{\kappa_{e,\textrm{ex}}}\delta\hat{a}_{e,\textrm{ex}} + \sqrt{\kappa_{e,\textrm{in}}}\delta\hat{a}_{e,\textrm{in}},\label{eq:microwave_eqn}
\\
& \dot{\hat{a}}_r = -i \Delta_r \hat{a}_r -\frac{\kappa_r}{2} \hat{a}_s -iJ \hat{a}_s  + \sqrt{\kappa_{r}}\delta\hat{a}_{r,\textrm{in}}. \label{eq:tm_mode_eqn}
\end{eqnarray}
\end{subequations}
Here, $\kappa_j$, $\kappa_{j,\textrm{ex}}$ and $\kappa_{j,\textrm{in}}$ are the total, extrinsic and intrinsic loss rates of respective modes, $\Delta_j$ are the detuning of the mode anhilation operators from their respective resonance frequencies, $\bar{F}_j$ are the coherent drive terms given by $|\bar{F}_j| = \sqrt{P_j/\hbar\omega_j}$ and $\delta\hat{a}_{j,\textrm{in}}$ and $\delta\hat{a}_{j,\textrm{ex}}$ represent the Langevin noise operators for bath and waveguide respectively. We take into account the finite mode matching between the free space coupled Gaussian mode of the single mode fiber and the evanescent field of the WGMR modes by multiplying $\kappa_{o,\textrm{ex}}$ with the mode amplitude overlap $\Lambda$ \cite{rueda_sanchez_resonant_2018}.
The noise operators follow the following correlations 
\begin{subequations}
\label{eq:noise_statistics} 
\begin{eqnarray}
& \braket{\delta\hat{a}_{j,k}(t) \delta\hat{a}_{j,k}(t')^\dagger} = (\bar{n}_k + 1)\delta(t-t')  ,\label{eq:noise_statistics1}
\\
& \braket{\delta\hat{a}_{j,k}(t)^\dagger \delta\hat{a}_{j,k}(t')} = \bar{n}_k \delta(t-t')  ,\label{eq:noise_statistics2}
\end{eqnarray}
\end{subequations}
where $k\in (\textrm{in}, \textrm{ex})$.
For optics, both $\bar{n}_\textrm{in}=0$ and $\bar{n}_\textrm{ex}=0$, while for microwave, $\bar{n}_\textrm{in}=N_b$ and $\bar{n}_\textrm{ex}=N_\textrm{wg}$.

\subsection{Coherent time-domain dynamics}
We first focus on the coherent time domain dynamics in our quantum transducer. By linearizing the intra-cavity field for the modes $\hat{a}_j = \bar{a}_j + \delta \hat{a}_j$, with $\bar{a}_j$ being the coherent field amplitude and $\delta \hat{a}_j$ the field fluctuations, we obtain the following coherent dynamics for the optical and microwave modes:
\begin{subequations}
\label{eq:timederivatives_coherent} 
\begin{eqnarray}
& \dot{\bar{a}}_p = -i \Delta_p \bar{a}_p -\frac{\kappa_o}{2} \bar{a}_p   -ig_0 (\bar{a}_s\bar{a}_e + \bar{a}_e^* \bar{a}_o) + \Lambda\sqrt{\kappa_{o,\textrm{ex}}}\bar{F}_p ,\label{eq:pump_eqn}
\\
& \dot{\bar{a}}_o = -i \Delta_o \bar{a}_o -\frac{\kappa_o}{2} \bar{a}_o -ig_0 \bar{a}_p\bar{a}_e  + \Lambda\sqrt{\kappa_{o,\textrm{ex}}}\bar{F}_o ,\label{eq:plus_fsr_eqn}
\\
& \dot{\bar{a}}_s = -i \Delta_s \bar{a}_s -\frac{\kappa_o}{2} \bar{a}_s -ig_0 \bar{a}_p\bar{a}_e^*  - i J \bar{a}_e  ,\label{eq:minus_fsr_eqn}
\\
& \dot{\bar{a}}_e = -i \Delta_e \bar{a}_e -\frac{\kappa_e}{2} \bar{a}_e -ig_0 \bar{a}_p\bar{a}_s^* -ig_0 \bar{a}_p^* \bar{a}_o  + \sqrt{\kappa_{e,\textrm{ex}}}\bar{F}_e ,\label{eq:microwave_eqn}
\\
& \dot{\bar{a}}_r = -i \Delta_r \bar{a}_r -\frac{\kappa_r}{2} \bar{a}_r -iJ \bar{a}_s.   \label{eq:tm_mode_eqn}
\end{eqnarray}
\end{subequations}

The output fields are then calculated as $\bar{a}_{e,\textrm{out}} = \sqrt{\kappa_{e,\textrm{ex}}}\bar{a}_e - \bar{F}_e$ for the microwave mode and $\bar{a}_{o,\textrm{out}} = \Lambda\sqrt{\kappa_{o,\textrm{ex}}}\bar{a}_o - \bar{F}_o$ for the upconverted optics mode. 
The above system of equations are numerically solved in the time domain using Euler's method. We use this method to model the classical time-dynamics of the system with arbitrary coherent pump pulses, thereby, fitting the experimental results shown in Figs.~1 and 2 of main text.
%%%%%%%%%%%%%%%%%%%%%%%%%%%
%%%%%%%%%%%%%%%%%%%%%%%%%%%
%%%%%%%%%%%%%%%%%%%%%%%%%%%

\subsection{Steady state model}
%For the analysis in steady state we turn to 
%the matrix formalism of Eqs.~\ref{eq:timederivatives2} and 
%Fourier space. 
After linearizing Eqs.~\ref{eq:timederivatives2}, valid for a strong classical pump tone, we rewrite this equation set in matrix form as
\begin{equation}
\dot{\boldsymbol{v}}(t) = \boldsymbol{M} \boldsymbol{v}(t) + \boldsymbol{K}\boldsymbol{A}(t),
\label{eq:matrix_eqn}
\end{equation}
where $\boldsymbol{v}(t)$ is $[\delta\hat{a}_o, \delta\hat{a}_o^\dagger, \delta \hat{a}_e, \delta \hat{a}_e^\dagger, \delta \hat{a}_s, \delta \hat{a}_s^\dagger,  \delta\hat{a}_r, \delta \hat{a}_r^
\dagger]^T$,
\begin{equation}
\boldsymbol{M} =
\begin{bmatrix}
-i\Delta_o - \frac{\kappa_o}{2} & 0 &  -ig & 0 & 0 & 0 & 0 & 0\\
0 & i\Delta_o - \frac{\kappa_o}{2} & 0 &  ig^* & 0 & 0 & 0 & 0 \\
-ig^* & 0 & -i\Delta_e - \frac{\kappa_e}{2} & 0 & 0 & -ig & 0 & 0 \\
0 & ig & 0 & i\Delta_e - \frac{\kappa_e}{2}  & ig^* & 0 & 0 & 0 \\
0 & 0 & 0 & -ig  & -i\Delta_s - \frac{\kappa_s}{2}& 0 & -iJ & 0 \\
0 & 0 & ig^* & 0 & 0  & i\Delta_s - \frac{\kappa_s}{2} & 0 & iJ \\
0 & 0 & 0 & 0    & -iJ & 0 & -i\Delta_r - \frac{\kappa_r}{2} & 0 \\
0 & 0 & 0 & 0    & 0 & iJ & 0 & i\Delta_r - \frac{\kappa_r}{2}
\end{bmatrix}
,
\end{equation}

\begin{equation}\setlength\arraycolsep{4pt}
\boldsymbol{K} = 
\begin{bmatrix}
\sqrt{\kappa_{o,\textrm{in}}} & 0 & \sqrt{\kappa_{o,\textrm{ex}}} & 0       & 0 & 0 & 0 & 0   & 0 & 0 & 0 & 0    & 0 & 0 \\ 
0 & \sqrt{\kappa_{o,\textrm{in}}} & 0 &  \sqrt{\kappa_{o,\textrm{ex}}}    & 0 & 0 & 0 & 0   & 0 & 0 & 0 & 0    & 0 & 0 \\ 
0 & 0 & 0 & 0  & \sqrt{\kappa_{e,\textrm{in}}} & 0 & \sqrt{\kappa_{e,\textrm{ex}}} & 0     & 0 & 0 & 0 & 0    & 0 & 0 \\ 
0 & 0 & 0 & 0  & 0 & \sqrt{\kappa_{e,\textrm{in}}} & 0 & \sqrt{\kappa_{e,\textrm{ex}}}     & 0 & 0 & 0 & 0    & 0 & 0 \\
0 & 0 & 0 & 0  &    0 & 0 & 0 & 0  &  \sqrt{\kappa_{s,\textrm{in}}} & 0 & \sqrt{\kappa_{s,\textrm{ex}}} & 0   & 0 & 0 \\
0 & 0 & 0 & 0  &    0 & 0 & 0 & 0  &   0 & \sqrt{\kappa_{s,\textrm{in}}} & 0 & \sqrt{\kappa_{s,\textrm{ex}}}  & 0 & 0 \\
0 & 0 & 0 & 0  &    0 & 0 & 0 & 0  &   0 & 0 & 0 & 0  &  \sqrt{{\kappa_r}} & 0 & \\
0 & 0 & 0 & 0  &    0 & 0 & 0 & 0  &   0 & 0 & 0 & 0  &  0 & \sqrt{\kappa_{r}} 
\end{bmatrix}
,
\end{equation}
and $\boldsymbol{A}(t) = [\delta\hat{a}_{o,\textrm{in}}, \delta\hat{a}_{o,\textrm{in}}^\dagger, \delta\hat{a}_{o,\textrm{ex}}, \delta\hat{a}_{o,\textrm{ex}}^\dagger, \delta\hat{a}_{e,\textrm{in}}, \delta\hat{a}_{e,\textrm{in}}^\dagger, \delta\hat{a}_{e,\textrm{ex}}, \delta\hat{a}_{e,\textrm{ex}}^\dagger, \delta\hat{a}_{s,\textrm{in}}, \delta\hat{a}_{s,\textrm{in}}^\dagger, \delta\hat{a}_{s,\textrm{ex}}, \delta\hat{a}_{s,\textrm{ex}}^\dagger, \delta\hat{a}_{r}, \delta\hat{a}_{r}^\dagger]$. Here, $g = g_0\sqrt{\bar{n}_p}$ is the parametrically enhanced electro-optic coupling strength and $\bar{n}_p$ the intra-cavity optical pump photon number. 
Equation~\ref{eq:matrix_eqn} is solved in the Fourier domain, yielding
\begin{equation}
\boldsymbol{v} (\omega) = \boldsymbol{S}(\omega) \boldsymbol{A}'(\omega),
\end{equation}
where $\boldsymbol{S} = [-\boldsymbol{M} - i\omega\boldsymbol{\mathds{1}}]^{-1}$ and $\boldsymbol{A}'(\omega) = \boldsymbol{K}\boldsymbol{A}(\omega)$. 
The output field can be obtained via the input-output theorem \cite{gardiner_input_1985,tsang_cavity_2011},
\begin{equation}
\delta \hat{a}_{j,\textrm{out}} (\omega) = - \delta \hat{a}_{j,\textrm{in}} +\sqrt{\kappa_{j,\textrm{ex}}} \delta \hat{a}_{j},
\end{equation}
with \textit{j=o,s}.
The total conversion efficiency on resonance is calculated from a single matrix element as
\begin{equation}
\eta_\textrm{tot} = |S_{oe}|^2 = |S_{eo}|^2 = \Lambda^2 \eta_e \eta_o \frac{4C\left(1+C_J^{-1}\right)^2}{\left(1+C+C_J^{-1}\right)^2},
\label{eq:conversion_eff}
\end{equation}
where $\eta_j = \kappa_{j,\textrm{ex}}/\kappa_j$ is the mode coupling efficiency, $C$ is the multi-photon cooperativity defined as $C=4\bar{n}_p g_0^2/(\kappa_o\kappa_e)$ and $C_J$, similarly, is the cooperativity related to the coupling $J$ defined as $C_J = 4J^2/(\kappa_o\kappa_r)$.
The factor $\Lambda^2$ is introduced by rewriting the external optical linewidth $\kappa_{o,\textrm{ex}} \rightarrow \Lambda^2 \kappa_{o,\textrm{ex}}$.

Note, for high $C_J$, the avoided crossing in the lower FSR optical mode is fully split, resulting in perfect suppression and no participation of the lower frequency sideband mode. Hence, in this limit, the above formula for conversion efficiency reduces to the usual two mode model, $\eta_\textrm{2-mode} = 4\Lambda^2 \eta_e \eta_o {C}/{\left(1+C\right)^2}$\cite{hease_bidirectional_2020}. Furthermore, for $C\ll1$, we get back the linear dependence of conversion efficiency on $C$ and no dependence on the value of $C_J$. The opposite limit is $C_J=0$ which means there is no avoided crossing in lower FSR optical mode and, thus, equal participation of both optical modes on either side of optical pump. In this limit, we get the maximum possible gain and the conversion efficiency formula reduces to $\eta_\textrm{J=0} = 4 \Lambda^2 \eta_e \eta_o {C}$. In contrast to the case of just two modes, here the conversion efficiency does not saturate as $C$ approaches 1. 

\subsection{Noise analysis}
%Next, we move on to noise analysis in steady state.
The noise spectrum of the output field can be obtained as $S_{jj,\mathrm{out}}(\Omega) = \int_{-\infty}^{+\infty} \left<\delta \hat{a}^{\dagger}_{j,\mathrm{out}}(t)\delta \hat{a}_{j,\mathrm{out}}(t^{\prime})\right> e^{i \Omega t} dt$, via the Wiener-Khinchin theorem.
The full expression of $S_{jj,\mathrm{out}}(\Omega)$ is too long to show here. This is due to the complex mode coupling scheme and because the optical and microwave cavity linewidths are comparable in magnitude.
The output photon flux is obtained by integrating the output noise spectrum over the specific measurement bandwidth
\begin{equation}
    N^j_{\mathrm{out}} =
    \int^{+\infty}_{-\infty} \chi(\Omega) S_{jj,\mathrm{out}}(\Omega) d\Omega,
\end{equation}
where $\chi(\Omega)$ is the measurement filter function. In our experiment, we use a Gaussian filter with full width half max (FWHM) of \SI{10}{MHz}.
The equivalent input noise is calculated accordingly
\begin{equation}
    N^j_{\mathrm{in}} =N^j_{\mathrm{out}}/\eta_\textrm{tot}.
\end{equation}
The predicted output noise takes gain into account and we use the total conversion efficiency $\eta_\textrm{tot}$ 
that also includes finite gain and coupling losses 
in order to infer the equivalent added noise referenced to the input port where the (quantum) signal is fed into the transducer.

In the absence of the optical pump (C=0), the output microwave noise spectrum is simply given as \cite{hease_bidirectional_2020},
\begin{equation}
N_{\mathrm{out}, C=0}^e (\Omega) = \frac{4 \kappa_{e,\mathrm{in}} \kappa_{e,\mathrm{ex}} } {\kappa_e^2  + 4\Omega^2} (N_b- N_{wg}) + N_{wg}.
\end{equation}
We use the output noise in absence of the optical pump to infer the bath temperature and subsequently calculate the equivalent mode occupancy of the microwave mode
\begin{equation}
    N_e = \frac{N_b \kappa_{e,\mathrm{in}}+ N_{wg} \kappa_{e,\mathrm{ex}}}{\kappa_{e,\mathrm{in}} + \kappa_{e,\mathrm{ex}} }.
\end{equation}
The calibration of the added noise during conversion critically depends on the measurement apparatus, which is discussed in detail in following sections. 

%%%%%%%%%%%%%%%%%%%%%%%%%%%%%%%%%%%%%%%%%%%%%%%%%%%
%%%%%%%%%%%%%%%%%%%%%%%%%%%%%%%%%%%%%%%%%%%%%%%%%%%
%%%%%%%%%%%%%%%%%%%%%%%%%%%%%%%%%%%%%%%%%%%%%%%%%%%

\section{Experimental Setup}\label{section:Setup}
\begin{figure}[t!]
	\centering
		\includegraphics[width=0.8\linewidth]{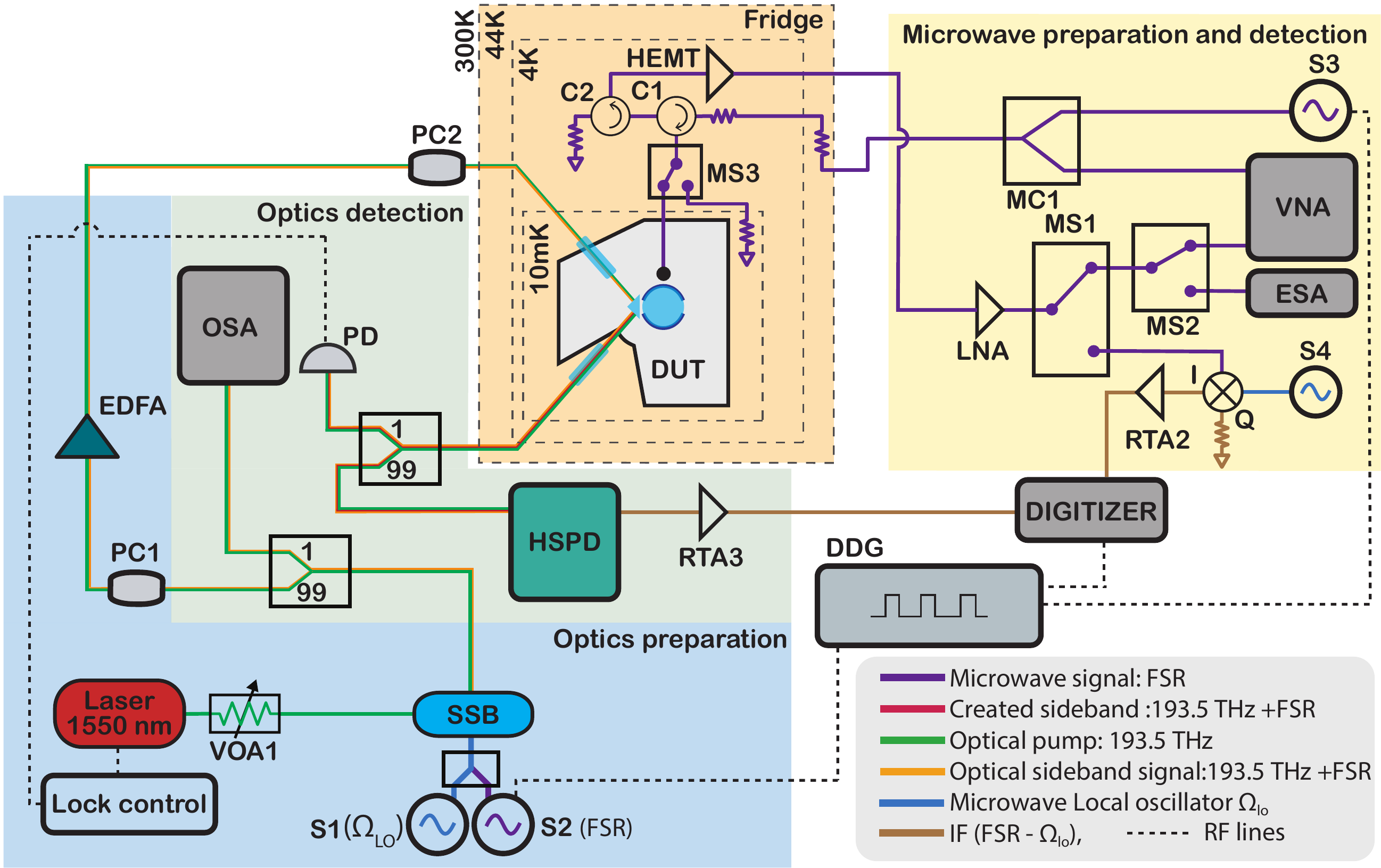}
 	\caption{\textbf{Experimental setup for low cooperativity measurements.} A tunable laser at frequency $\omega_p$ is sent through a variable optical attenuator (VOA1) to control the power output. Thereafter, the laser is sent to a single sideband (SSB) modulator. The modulator operates in a mode where the central pump frequency is allowed to pass through and only the lower sideband is suppressed. The SSB is connected to two microwave sources. The arrangement allows us to independently control the optical signal tone and the optical local oscillator (LO) tone, \SI{200}{MHz} detuned, independently. The optical signal source S2 is also connected to a digital delay generator (DDG) to make pulses at the right time. The modulated output from the SSB is divided in two parts - 1\% is used to monitor the suppression ratio of the sidebands using an optical spectrum analyzer (OSA), 99\% is sent to an Erbium-doped fiber amplifier (EDFA) and amplified before being sent to the dilution refrigerator (DR). In the DR, the light is focused via a grin lens on the surface of the prism and coupled to the optical whispering gallery mode resonator (WGMR) via evanescent coupling. Polarization controllers PC1 and PC2 are adjusted to efficiently couple to the TE modes of optical WGMR. The output light from the optical WGMR is sent in a similar fashion into the collection grin lens. Outside the DR, the output light is separated in two parts - 1\% is detected directly on a photo-diode (PD) to lock the laser to the optical pump mode and 99\% is sent to a high speed photo-diode HSPD (\SI{400}{MHz}). The presence of the optical LO and signal in the same fiber means that the optical signal can be easily detected at the set frequency via downconversion. The output signal from HSPD is sent to an amplifier RTA3 before sent for digitization. On the microwave side, the signal is sent from the microwave source S3 which is also connected to DDG for accurately timed pulse generation (or from the VNA for microwave mode spectroscopy) to the fridge input line via the microwave combiner (MC1). The input line is attenuated with attenuators distributed between \SI{3}{K} and \SI{10}{mK} with a total of \SI{60}{dB} in order to suppress room temperature microwave noise. Circulator C1 redirects the reflected tone from the cavity to the amplified output line, while C2 redirects noise coming in from the output line to a matched \SI{50}{\ohm} termination. The output line is amplified at \SI{3}{K} by a HEMT-amplifier and then at room temperature again with a low noise amplifier (LNA). The output line is connected to switch MS1 and MS2, to select between an ESA, a VNA or a digitizer measurement via manual downconversion using MW LO S4 (\SI{200}{MHz} detuned). Lastly, microwave switch MS3 allows to swap the device under test (DUT) for a temperature $T_{\textrm{\SI{50}{\ohm}}}$ controllable load, which serves as a broad band noise source in order to calibrate the output line's total gain and added noise \cite{hease_bidirectional_2020}.}
	\label{fig:setup_low_coop}
\end{figure}

\begin{figure}[t!]
	\centering
		\includegraphics[width=0.8\linewidth]{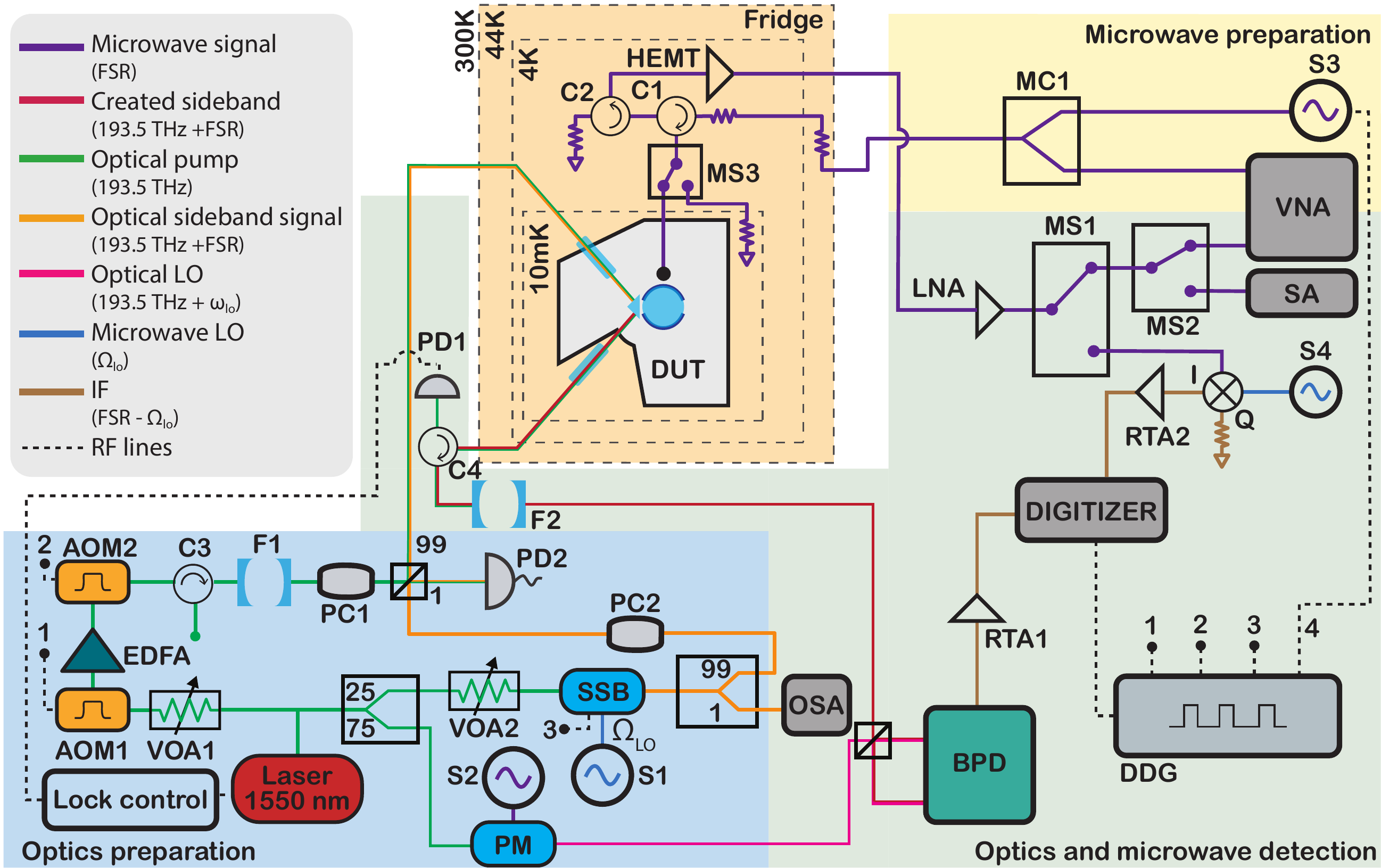}
 	\caption{\textbf{Experimental setup for high cooperativity measurement.} A tunable laser at frequency $\omega_p$ is divided into two equal parts - one to serve as the optical pump and the other to produce the optical signal and the optical local oscillator (LO). The optical pump side (left) first passes through a variable optical attenuator (VOA1) to control the power sent in this arm and is then sent to an acousto-optic modulator (AOM1). The AOM is accurately pulsed using a digital delay generator (DDG). The produced optical pulsed are sent to an Erbium-doped fiber amplifier (EDFA) where they are amplified. The output of the EDFA is sent to another AOM2. The second AOM is also connected to DDG and acts as a window filter in time to suppress the broad band spontaneous emission noise from the EDFA. The optical pump pulse is further cleaned with a filter cavity F1, which is locked to the laser frequency (circulator C3 ensures that reflected noise from cavity F1 is dissipated) before being combined with the signal arm and sent to the dilution refrigerator (DR). The signal arm (right of the laser) is first divided in two parts - the optical LO and the optical signal. 25\% of the light in signal arm is used to produce the optical signal. After passing through attenuator VOA2, the optical signal is produced using a single sideband modulator (SSB). This time we suppress both the central pump frequency and the lower sideband frequency keeping only the upper sideband as the optical signal. The SSB is driven by a microwave source S1 which is also connected to the DDG to accurately pulse the optical signal. The optical signal is divided in two parts - 1\% is reserved to monitor the sideband suppression ratio via an optical spectrum analyzer (OSA), 99\% is sent to DR after combining with the optical pump. The 75\% light on the right side of the laser is used to produce the optical LO via a phase modulator (PM). The PM is operated via a microwave source S2 with a power such that the central tone is suppressed. Finally, the optical LO is sent directly to the optical heterodyne setup. The optics inside the DR has been explained in the caption of Fig.~\ref{fig:setup_low_coop}. The output light from the DR is sent to filter F2 to reject the strong optical pump. The reflected optical pump from cavity F2 is captured by photodiode PD1 via the circulator C4. The reflected optical pump measurement is used to lock the laser to the optical pump mode. The cleaned optical signal is sent to the heterodyne setup and measured with a balanced photo-detector (BPD). The output signal is amplified via a room temperature amplifier RTA1 before being sent to a digitizer. The microwave side of the setup is explained in the caption of Fig.~\ref{fig:setup_low_coop}. }
	\label{fig:setup_high_coop}
\end{figure}

%We conducted the experiment in two parts - one for low cooperativities and one for high cooperativities. 
The experimental setups for low and high cooperativity conversion were slightly different and optimized for each measurement. The setup for low (high) cooperativity measurements is shown and described in Fig.~\ref{fig:setup_low_coop} (Fig.~\ref{fig:setup_high_coop}). The main difference between the setups is whether the optics signal and the optics LO are separated. In case of the low cooperativity setup, we keep them always in the same optical fiber for better phase stability. In case of the high cooperativity setup we prepare the high power optical pump separately before combining it with the optical signal. Later, the optical signal is separated from the optical pump using an optical filter. 

\section{System Characterization} 
Our experiment is divided in two broad parts - measurements at low cooperativity ($<10^{-3}$) and measurements at high cooperativity ($\sim 1$). We use a different set of optical modes for these experiments because of a large time gap of a few months between the experiments. In the following subsections, we detail the system characterization done for both parts of the experiment. Since we use the same device for this experiment, the electro-optic coupling rate $g_0$ is the same as reported in our previous manuscript \cite{hease_bidirectional_2020}.

\subsection{Low cooperativity characterization}
Before doing conversion measurements, we independently characterize the optical and microwave modes. We characterize these modes in both the time and frequency domain. Characterization of the optics mode in frequency domain is described in detail in reference \cite{hease_bidirectional_2020}. We move the coupling prism via a piezo positioner, to change the external coupling rate $\kappa_{o,\textrm{ex}}$, and fit the mode profile in frequency domain to extract the linewidth. The normalized mode spectrum is given as
\begin{equation}
\frac{\left|S_{o o}\left(\omega-\omega_{o}\right)\right|^{2}}{\left|\mathrm{~S}_{o o}(\Delta \omega)\right|^{2}}=1-\frac{4 \kappa_{o, \mathrm{ex}} \Lambda^{2}\left(\kappa_{o}-\Lambda^{2} \kappa_{o, \mathrm{ex}}\right)}{\kappa_{o}^{2}+4\left(\omega-\omega_{o}\right)^{2}},
\end{equation} 
where $\omega_o$ is the optical mode resonance frequency. Fitting the measured optical spectrum, we determine the internal linewidth, coupling efficiency and mode matching factor, $\Lambda$, for the optical mode. 
Similarly, we also fit the microwave mode spectrum measured via a Vector Network Analyzer (VNA) to determine the total linewidth of microwave mode and its coupling efficiency. We then use the parameters obtained from frequency domain characterization to verify our time-domain characterization described next.

\begin{figure}[ht!]
	\centering
		\includegraphics[scale=0.45]{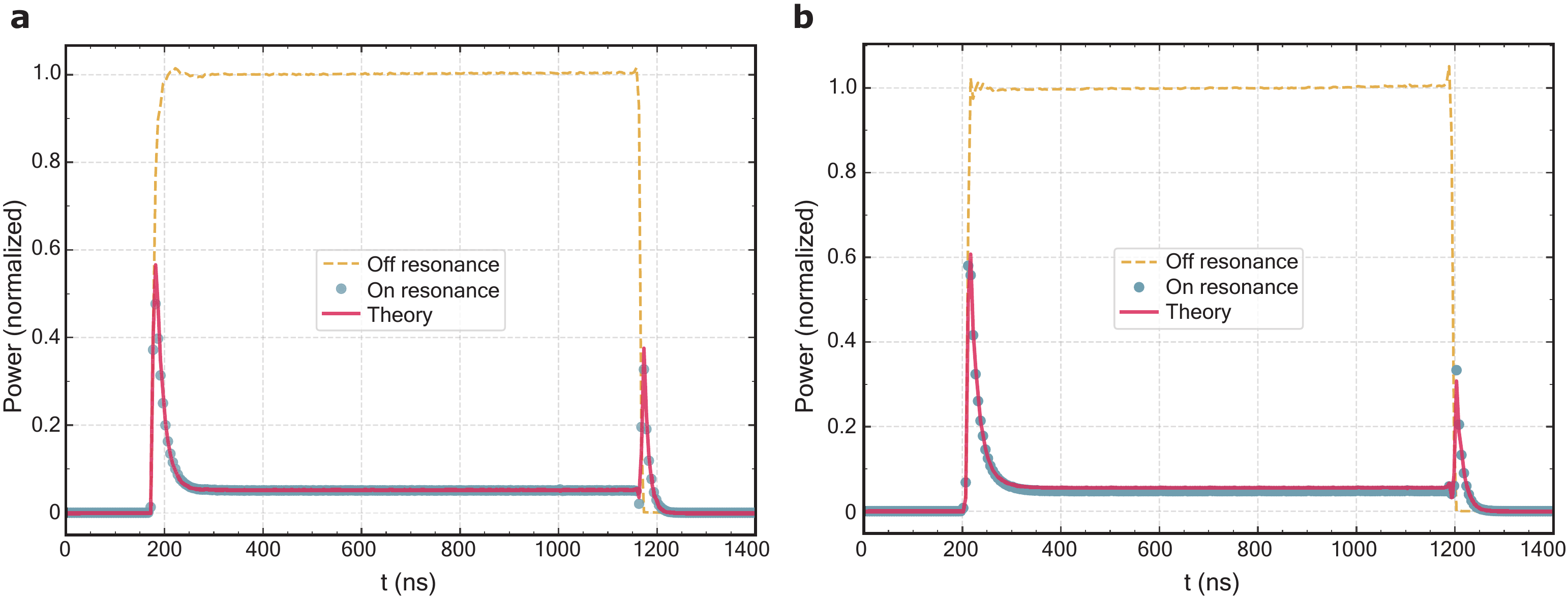}
	\caption{\textbf{Time domain system characterization for low cooperativity measurements.} 
\textbf{a} (\textbf{b}), Normalized pulse reflection of a square pulse from the optical (microwave) cavity on and off resonance. The red curve is a theoretical fit obtained using time domain input-output theory.}
	\label{fig:low_coop_refl}
\end{figure}

Figure \ref{fig:low_coop_refl} shows the time domain characterization of the optics and microwave modes. We characterize the system by sending a square pulse of a coherent tone to the cavity on and off resonance. The off resonant pulse reflects without any modifications and, thus, is used as a measurement of the input pulse shape. We use Eqs.~\ref{eq:plus_fsr_eqn} and \ref{eq:microwave_eqn} as well as the input pulse shape to solve for modes $\bar{a}_o$ and $\bar{a}_e$ and, subsequently, calculate $\bar{a}_{o, \textrm{out}}$ and $\bar{a}_{e, \textrm{out}}$ as detailed in section \ref{sec:theory} to predict the pulse reflection on resonance and, finally, calculate the power reflected on resonance, $|\bar{a}_{j, \textrm{out}}|^2$. 
This prediction is used to fit the reflected power in Fig.~\ref{fig:low_coop_refl}. The fitted system parameters are shown in Table~\ref{tab:low_coop}. The parameters fully agree with those determined by frequency domain characterization except for the optical mode matching factor, $\Lambda$, which was determined to be \num{0.806} from the frequency domain characterization and \num{0.838} from the time domain characterization.
We use the value obtained from time domain characterization which we believe is more accurate and directly applicable to the time domain conversion measurements.

\newcolumntype{M}[1]{>{\centering\arraybackslash}m{#1}}
\begin{table}
\caption{\label{tab:low_coop}System parameters for low cooperativity measurements.}
\begin{ruledtabular}
\begin{tabular}{|M{4cm}|M{7cm}|M{4cm}|}
Parameter & Description & Value\\
\hline
$\omega_o/2\pi$ & Optical signal frequency & \SI{193}{THz}\\
$\kappa_o/2\pi$ & Optical signal linewidth & \SI{15.55}{MHz}\\
$\eta_o$ & Optical coupling efficiency & \num{0.55} \\
$\Lambda$ & Optical mode mismatch factor & \num{0.838} \\
$\omega_e/2\pi$ & Microwave mode frequency & \SI{8.803}{GHz}\\
$\kappa_e/2\pi$ & Microwave signal linewidth & \SI{12.12}{MHz}\\
$\eta_e$ & Microwave coupling efficiency & \num{0.369} \\
$g_0/2\pi$ & Electro-optic coupling rate & \SI{37}{Hz}\\
\end{tabular}
\end{ruledtabular}
\end{table}

The shape of reflected pulse on resonance in Fig.~\ref{fig:low_coop_refl} is interesting to understand in more detail. 
%The sharp peaks in the beginning and at the end of the pulse are because of finite rise time of the input pulse. 
The first peak occurs due to the rapid rise time of the input pulse which has much higher bandwidth than the cavity. Thus, all of it gets reflected before the cavity has the chance to absorb part of the input tone. The initial rise is interrupted as the cavity gets the time to absorb the input light and start re-emitting that light to cancel the input pulse reflection. This continues until a steady state is reached. 
%The steady state continues for the length of the pulse and is only disrupted at the end of pulse. 
At the end of the pulse the pulse power drops much faster than the cavity bandwidth. As a result, photon emission from the cavity does not get time to change. As the input pulse drops, there is a moment when the cavity emission perfectly cancels out the input reflection leading to a point of zero reflection (more clearly seen for shorter pulses, see Fig.~\ref{fig:high_coop_refl}). As the input pulse drops further, the emission from the cavity takes over and the reflected power rises again. Finally, only the emission from the cavity is left which slowly decays with cavity linewidth. 

For low cooperativities, we do not characterize the $\hat{a}_s$ mode hybridization since the conversion efficiency is approximately reduced to just the simple 3-mode beam-splitter like interaction where the TM coupling is irrelevant (see Eq.~\ref{eq:conversion_eff}).

\subsection{High cooperativity characterization}
\begin{figure}[ht!]
	\centering
		\includegraphics[width=\linewidth]{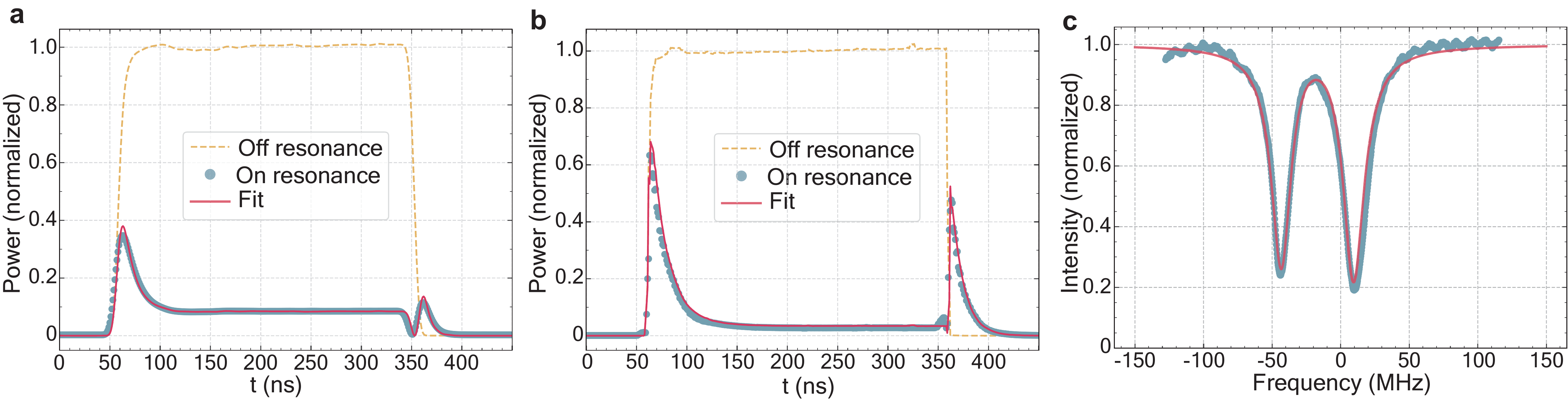}
	\caption{\textbf{System characterization for high cooperativity measurements.} 
 \textbf{a} (\textbf{b}), Normalized pulse reflection of a square pulse from the optical (microwave) cavity on and off resonance. The red curve is a theoretical fit obtained using time domain input-output theory. 
 \textbf{c}, Measured reflection spectrum around the red detuned TE optical mode $\hat{a}_s$. The mode shows an avoided crossing that is slightly detuned from the FSR at zero frequency because of its coupling to the near degenerate TM optical mode $\hat{a}_r$. The red line shows a theoretical fit of the split mode using Eq.~\ref{eq:splitting_spectrum}. }
	\label{fig:high_coop_refl}
\end{figure}

Figure \ref{fig:high_coop_refl}(a) and (b) shows the time domain characterization of the optical and microwave modes valid for high cooperativity measurements. The resulting system parameters obtained are shown in Table~\ref{tab:high_coop}. 
%For the microwave mode the parameters match perfectly with those obtained using frequency domain measurements. Steady state frequency domain measurements of the optical mode were not performed due to the prism being stuck... 
%However, due to experimental challenges, we were not able to repeat the frequency domain characterization for the optical modes in this case. 
Figure \ref{fig:high_coop_refl}(c) shows the optical Stokes mode $\hat{a}_s$ in the frequency domain. The mode is coupled to a degenerate TM optical mode $\hat{a}_r$, which results in an avoided crossing of the TE mode. We determine the strength of coupling between these modes, the total linewidth of the TM optical mode and its detuning by fitting the measured spectrum in the frequency domain. 

The frequency response of the split mode is derived by considering two coupled modes where only one of them is pumped
\begin{subequations}
\label{eq:coupled_modes} 
\begin{eqnarray}
& \dot{\bar{a}}_s = -i \Delta_s \bar{a}_s -\frac{\kappa_o}{2} \bar{a}_s  - i J \bar{a}_r  +  \Lambda\sqrt{\kappa_{o,\textrm{ex}}}\bar{F}_s  ,\label{eq:coupled_te_mode}
\\
& \dot{\bar{a}}_r = -i \Delta_r \bar{a}_r -\frac{\kappa_r}{2} \bar{a}_r -iJ \bar{a}_s.   \label{eq:coupled_tm_mode}
\end{eqnarray}
\end{subequations}
Solving the above set of equations in frequency domain, we find
\begin{equation}
a_s(\omega) = \frac{\Lambda\sqrt{\kappa_{o,\textrm{ex}}}\hat{F}_s}{(i \Delta_s + \kappa_s/2) + \frac{J^2}{i \Delta_r + \kappa_r/2}},
\label{eq:splitting_spectrum}
\end{equation}
where $\Delta_j =\omega_j - \omega $. Subsequently, we calculate the output field as $\bar{a}_{s, \textrm{out}}(\omega) = \Lambda\sqrt{\kappa_{o,\textrm{ex}}} \bar{a}_s - \bar{F}_s$. We use this formalism to fit the split mode in the frequency domain. The resulting parameters are reported in Table~\ref{tab:high_coop}.

\newcolumntype{M}[1]{>{\centering\arraybackslash}m{#1}}
\begin{table}
\caption{\label{tab:high_coop}System parameters for high cooperativity.}
\begin{ruledtabular}
\begin{tabular}{|M{4cm}|M{7cm}|M{4cm}|}
Parameter & Description  & Value\\
\hline
$\omega_o/2\pi$ & Optical signal frequency & \SI{193}{THz}\\
$\kappa_o/2\pi$ & Optical signal linewidth & \SI{25.8}{MHz}\\
$\eta_o$ & Optical coupling efficiency & \num{0.58} \\
$\Lambda$ & Optical mode mismatch factor & \num{0.78} \\
$\omega_e/2\pi$ & Microwave mode frequency & \SI{8.795}{GHz}\\
$\kappa_e/2\pi$ & Microwave signal linewidth & \SI{13.706}{MHz}\\
$\eta_e$ & Microwave coupling efficiency & \num{0.408} \\
$g_0/2\pi$ & Electro-optic coupling rate & \SI{37}{Hz}\\
$J/2\pi$ & TE-TM optical mode coupling rate & \SI{26.21}{MHz}\\
$\kappa_r/2\pi$ & Optical TM mode linewidth & \SI{9.96}{MHz} \\
$\Delta_s /2\pi$ & Optical TE mode detuning & \SI{15.5}{MHz} \\ 
$\Delta_r /2\pi$ & Optical TM mode detuning & \SI{19.5}{MHz} \\ 
\end{tabular}
\end{ruledtabular}
\end{table}

\section{4-port calibration}
The total conversion efficiency is determined from a four-port calibration procedure of the transducer \cite{andrews_bidirectional_2014,hease_bidirectional_2020}. For this purpose, the transducer is treated as a device with four ports - optics input (prism surface), optics output (prism surface), microwave input (coaxial coupler) and microwave output (same coaxial coupler) with optical (microwave) input transmission coefficient $\beta_1$ ($\beta_3$) of the setup as well as optical (microwave) output transmission coefficient $\beta_2$ ($\beta_4$) of the setup. %as shown in Fig.~\ref{fig:fourPort}\cite{hease_bidirectional_2020}. 
Then, by measuring the optical and microwave transmission along with the conversion efficiency in both direction, the total conversion efficiency is calculated as
\begin{equation}
\eta_{\textrm{tot}} = \sqrt{\frac{|S_{eo}(\omega_e)|^2 \cdot |S_{oe}(\omega_o)|^2}{|S_{oo}(\omega_{\Delta,o})|^2 \cdot |S_{ee}(\omega_{\Delta,e})|^2}}.
\end{equation}
Here, the conversion efficiencies $|S_{eo}|^2$ and $|S_{oe}|^2$ are measured on resonance, while the transmission $|S_{oo}|^2$ and $|S_{ee}|^2$ are measured off resonance ($\omega_{\Delta,o} \gg \kappa_o,\  \omega_{\Delta,e} \gg \kappa_e$). Knowing any one of the four gain/loss terms ($\beta_i$) in the transmission lines, we can calculate all the other $\beta_i$. In the present case we use the microwave gain $\beta_4$ as the known quantity, which was measured independently (see appendix of Ref.~\cite{hease_bidirectional_2020}) and calculate the remaining $\beta_i$.  We determine the following values: the optical input loss $\beta_1=$ \SI{-6.33}{dB}, the output optical gain $\beta_2=$ \SI{18.63}{dB}, the microwave input loss  $\beta_3=$ \SI{-74.92}{dB} and the microwave output gain $\beta_4=$ \SI{81.75}{dB}.

%\begin{figure}[ht!]
%	\centering
%		\includegraphics[scale=0.4]{fourPort}
% 	\caption{\textbf{Four port calibration} By measuring different $S_{ij}$ coefficients between the microwave and optical ports, we determine the different $\beta_i$. The optical input loss $\beta_1$ is \SI{-6.33}{dB} and the output optical gain $\beta_2$ (with heterodyne) is \SI{18.63}{dB}. The microwave input loss is $\beta_3$ is \SI{-74.92}{dB} and the microwave output gain $\beta_4$ is \SI{81.75}{dB}. }
%	\label{fig:fourPort}
%\end{figure}

\section{Optical heterodyne detection}
On the optical side, we 
%perform quantum limited 
detect the optical output signal using a balanced heterodyne setup, i.e. by beating the signal with a strong local oscillator with coherent optical field $\bar{a}_{\mathrm{LO}}$ at frequency of $\omega_o+\Omega_{\mathrm{LO}}$. This results in a balanced photocurrent 
$
\delta I(t) = i (-e^{i\Omega_{\textrm{LO}} t }  \bar{a}^{*}_{\textrm{LO}} \delta \hat{a}_{o,\textrm{out}} 
+ e^{-i\Omega_{\textrm{LO}} t } \bar{a}_{\textrm{LO}}\delta \hat{a}^{\dagger}_{o,\textrm{out}}) 
$. 
We thus obtain the symmetrized power spectral density 
$S_I(\Omega) =\frac{1}{2} \int_{-\infty}^{\infty}\langle\overline{\left\{\delta I\left(t+t^{\prime}\right), \delta I\left(t^{\prime}\right)\right\}}\rangle e^{i \Omega t} d t$.
The optical heterodyne efficiency or the equivalent noise floor level can be determined using the output gain $\beta_2$ and the absolute power measured in the baseline without any signal
\begin{equation}
P_{\textrm{baseline}} = \hbar \omega_o \beta_2 BW \bar{n}_{\mathrm{add}},
\end{equation}
where $BW$ the measurement bandwidth and $\bar{n}_{\textrm{add}}$ the equivalent noise in the heterodyne baseline. 

Using the output optical gain $\beta_2$, the equivalent noise floor in optics heterodyne $\bar{n}_{\textrm{add}}$ is calculated to be \num{34.3} photons. The optical detection efficiency is low in our case for a number of reasons. There is a  $\approx$ \SI{3}{dB} loss while coupling the light from the device output, i.e. from the prism surface,  to the optical fiber with a gradient index lens. An optical filter is used to reject the strong optical pump for the balanced heterodyne detection of the weak converted signals, which introduces another \SI{4}{dB} of loss. We use the first order sideband generated from a phase modulator as the optical LO. Since the phase modulator produces many other optical tones, it reduces the optical balanced heterodyne efficiency to about \SI{17}{\%}. These are technical nonidealities of the setup that can be improved in the future.

\section{Microwave noise measurements} 
On the microwave side we perform phase-insensitive amplification of the weak microwave signal using a cryogenic low-noise high-electron mobility transistor amplifier \cite{Clerk2010}. The amplified microwave field is sent through subsequent amplifiers, and  mixed with a microwave LO at room temperature. The calibrated added noise due to microwave detection chain is \num{12.74} photons\ \SI{}{s}$^{-1}$\SI{}{Hz}$^{-1}$, as shown in the appendix of Ref.~\cite{hease_bidirectional_2020}. In all of the noise measurements in the main text we report the measured added noise due to the transducer, i.e. with the constant added noise due to loss and amplifiers subtracted. 

In the main text we show microwave noise measurements with two different bandwidths - \SI{10}{MHz} for fast noise detection ($\sim$ \SI{100}{ns} time resolution) and \SI{100}{kHz} for slower noise detection but with better SNR. The main source of error for the slow, low occupancy measurements is the systematic absolute error due to long term drifts of the noise baseline that is subtracted. We measure and average the noise baseline over multiple hours (like the measurements) on subsequent days to determine the absolute standard error of $\pm0.02$ photons\ \SI{}{s}$^{-1}$\SI{}{Hz}$^{-1}$. This error along with the statistical error of the actual noise measurement is propagated to get the final error bars. 
In the case of fast measurements, which were averaged only over a few tens of minutes, we take into account a larger observed variation of the noise baseline on that timescale of $\pm0.1$ to $\pm0.2$ photons\ \SI{}{s}$^{-1}$\SI{}{Hz}$^{-1}$. 

%The \SI{100}{kHz} results for the average output microwave noise shown in Fig.~3(c) of the main text have been measured for slower repetition rates. For these measurements, we collect the statistics over three hours and report the standard error of the mean value also taking into account the baseline drift.

\begin{figure}[t]
	\centering
		\includegraphics[width=0.66\linewidth]{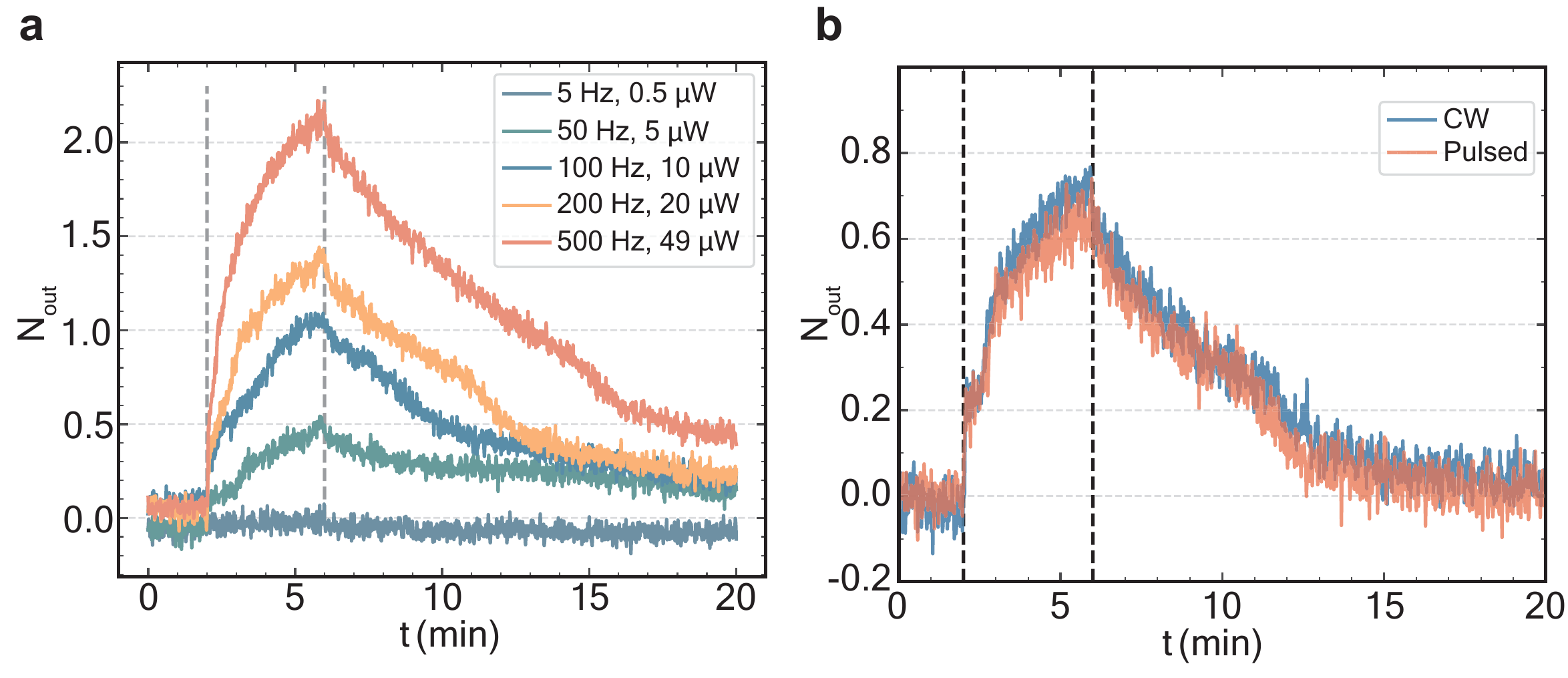}
 	\caption{\textbf{Low-bandwidth microwave noise measurements.} The output noise of the microwave cavity is measured with a \SI{100}{kHz} bandwidth. Each point is averaged for one second for better signal to noise ratio. \textbf{a}, Output microwave noise dynamics measured on long time scales as a function of pulse repetition rate for optical pump pulses corresponding to $C\sim 0.38$. The region between vertical dashed lines marks the time interval during which the optical pump pulses are turned on. \textbf{b}, Equivalence of added thermal microwave noise CW optical power and optical pulses with same average optical power.}
	\label{fig:slow_noise}
\end{figure}

The slow \SI{100}{kHz} measurements allow us
%do longer measurements 
to study noise dynamics of the system on a long time scale. Figure \ref{fig:slow_noise}(a) shows the average microwave noise output as the optical pump pulses corresponding to $C \sim 0.38$ with different repetition rates are turned on (marked region between vertical dashed lines). We observe that our system does not heat up immediately as soon as the pulses are turned on, rather it slowly reaches the steady state in a few minutes after the pulses are turned on.
Moreover, the cooling time is even longer and it can take up to an hour to come back to the equilibrium occupancy depending on the average optical pump power applied. 
%starting microwave occupancy. 
The specific time scales are expected to depend critically of the thermal contact and conductivity of the localized heat source (the dielectric-superconducting sample) to the cold bath (the mixing chamber plate), as well as on the cooling power of the dilution refrigerator \cite{mobassem2021}. 

Since this bulk electro-optic system has such long heating and cooling timescales, for short and fast repeating pulses, it is only the average optical power that determines the added thermal noise due to optical heating. Figure \ref{fig:slow_noise}(b) shows the equivalence between heating via optical pulses (red) and continuous optical power (blue) during the time interval marked by the two vertical dashed lines. 
%The red color shows the average microwave noise output when we turn on the pulses in the region marked by vertical dashed lines. 
We observe the same dynamics of the output microwave noise if the power level of average optical power is matched.  
%in the pulses including the wait time between pulses. 
This experiment shows that thermal noise can be tuned by changing the pulse repetition rates while keeping the same level of cooperativity. This feature of the transducer allows to see amplified vacuum noise with negligible thermal noise by decreasing the pulse repetition rate and maintaining the high cooperativity.

\section{Kerr Effect}
\begin{figure}[h!]
	\centering
		\includegraphics[scale=0.5]{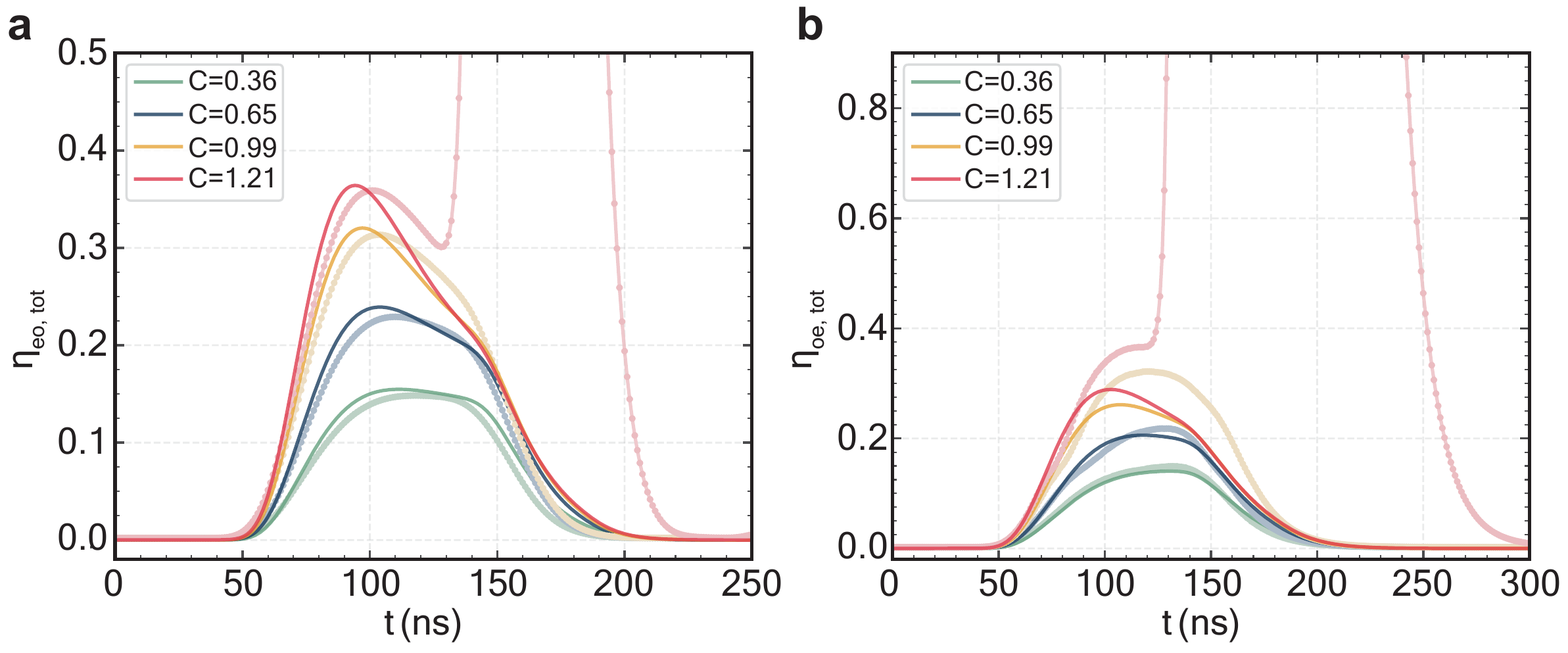}
 	\caption{\textbf{Parametric optical amplification due to the $\chi^{(3)}$ non-linearity.}  \textbf{a} (\textbf{b}), Time domain conversion in the microwave to optics (optics to microwave) direction for \SI{100}{ns} optical pump pulses of different power and $C$. The points joined by light colored lines are measured experimentally. The thin bright lines are theoretical only taking into account the $\chi^{(2)}$ effect. After $C\approx 1$, we observe a delayed parametric amplification event. The measured transmission reaches as high as 11.4 in case of microwave to optics conversion and 13.1 in case of microwave to optics conversion (not visible).}
	\label{fig:kerr_effect}
\end{figure}

In Fig.~2 of the main text we show conversion up to a cooperativity of \num{0.92}. This is because after a certain threshold input power, which also depends on the optical pump pulse length, we observe an extreme amplification at the end of the conversion pulse. Figure \ref{fig:kerr_effect} shows this effect for both microwave-to-optics and optics-to-microwave conversion cases. 

After a number of tests we came to the conclusion that this effect is most likely due to the third-order $\chi^{(3)}$ nonlinearity in lithium niobate, an effect that is commonly utilized in optical parametric amplifiers \cite{kippenberg_kerr-nonlinearity_2004}. We verified that amplification in the optical signal is present even when there is no coherent signal drive present and for cases when the microwave mode is far detuned from the optical FSR. This proves that the effect is independent of the usual $\chi^{(2)}$ nonlinear interaction. However, when the microwave mode is matched with the optical FSR, we see amplification in the microwave signal as well due to strong optical sideband combined the $\chi^{(2)}$ mediated beam splitter interaction (conversion of the amplified optical signal). Note that the cooperativity threshold for amplification of $C\approx 1$, as shown in Fig.~\ref{fig:kerr_effect}, is a coincidence and only valid for \SI{100}{ns} optical pump pulses. For longer optical pump pulses, the cooperativity threshold for amplification becomes smaller and well below unity. 

This parameter regime of seeing effects of both $\chi^{(2)}$ and $\chi^{(3)}$ nonlinearities together is, to best of our knowledge, novel. Producing coherent phase-locked microwave and optical drives together requires a systematic investigation and may prove useful in future. 

\end{document}